\documentclass[11pt]{article}
\usepackage{graphicx}
\usepackage{times}
\usepackage{amssymb}
\usepackage[numbers,sort,square,comma]{natbib}
\usepackage[vcentering,dvips]{geometry}
\def\gsim { \lower .75ex \hbox{$\sim$} \llap{\raise .27ex \hbox{$>$}} }
\def\lsim { \lower .75ex \hbox{$\sim$} \llap{\raise .27ex \hbox{$<$}} }

\newcommand{\apj}{ApJ}
\newcommand{\apjl}{ApJL}
\newcommand{\apjs}{ApJS}
\newcommand{\aj}{AJ}
\newcommand{\mnras}{MNRAS}
\newcommand{\nat}{Nature}

\newcommand{\araa}{ARA\&A}
\newcommand{\aapr}{A\&ARv}
\newcommand{\aap}{A\&A}

\newcommand{\pasp}{Astronomical Society of the Pacific}

\begin{document}

\title{The Cosmic History of Black Hole Growth from Deep Multiwavelength Surveys}

\author{
{Ezequiel Treister$^{1,2,3}$ and C. Megan Urry$^{4,5,6}$}
\\
\\
$^{1}$ Universidad de Concepci\'{o}n, Departamento de Astronom\'{\i}a, Casilla 160-C, Concepci\'{o}n, Chile\\
$^{2}$ Institute for Astronomy, 2680 Woodlawn Drive, University of Hawaii, Honolulu, HI 96822\\
$^{3}$ Chandra/Einstein Fellow\\
$^{4}$ Yale Center for Astronomy and Astrophysics, P.O. Box 208121, New Haven, CT 06520.\\
$^{5}$ Department of Physics, Yale University, P.O. Box 208121, New Haven, CT 06520.\\
$^{6}$ Department of Astronomy, Yale University, PO Box 208101, New Haven, CT 06520.\\
\\
Correspondence should be addressed to Ezequiel Treister, etreiste@astro-udec.cl
\\
}
\date{\today}
\maketitle

\begin{abstract}
Significant progress has been made in the last few years on understanding how supermassive black holes
form and grow. In this paper, we begin by reviewing the spectral signatures of Active Galactic 
Nuclei (AGN) ranging from radio to hard X-ray wavelengths. We then describe the most commonly used methods
to find these sources, including optical/UV, radio, infrared and X-ray emission and optical emission lines.
We then describe the main observational properties of the obscured and unobscured AGN population. Finally, we
summarize the cosmic history of black hole accretion, i.e., when in the history of the Universe supermassive black holes 
were getting most of their mass. We finish with a summary of open questions and a description of planned and future 
observatories that are going to help answer them.
\end{abstract}

\section{Introduction} 

Astrophysical black holes come in a wide range of masses, from $\gsim$3 $M_\odot$ for
stellar mass black holes \cite{orosz03} to $\sim$10$^{10}$ $M_\odot$ for so-called
supermassive black holes \cite{lynden-bell69,lauer07}. The best evidence for the existence of
a supermassive black hole can be found in the center of the Milky Way galaxy, where from dynamical studies, 
the mass of the Sgr A$^*$ source  was established to be $\sim$4.4$\times$10$^{6}$$M_\odot$ \cite{genzel97,ghez98}.

Evidence for the existence of supermassive black holes has also been found in other massive nearby 
galaxies \cite{kormendy95}, mostly from resolved stellar and gas kinematics. For active galaxies, it has been possible
to use the technique known as reverberation mapping \cite{bahcall72,blandford82,peterson93}. From these observations, 
a clear correlation has been established between the mass of the central black hole and properties of the host galaxy such 
as stellar mass in the spheroidal component \cite{marconi03}, luminosity \cite{magorrian98}, velocity dispersion 
\cite{ferrarese00,gebhardt00} and mass of the dark matter halo \cite{ferrarese02}. The fact that such correlations exist, 
even though these components have very different spatial scales, suggests a fundamental relationship between black hole
formation and galaxy evolution. Furthermore, it is now well established by simulations \cite{springel05} that the energy 
output from the growing central black hole can play a significant role in the star formation history of the host galaxy. In 
particular, theory suggests that nuclear activity regulates star formation either by removing all the gas 
\cite{hopkins06,menci06} or by heating it \cite{croton06}. It is therefore obvious, that a complete study of galaxy evolution 
requires a comprehensive understanding of black hole growth.

Most current black hole formation models tell us that the first black hole seeds formed at $z$$\gsim$15. While the exact
mechanism for the formation of the first black holes is not currently known, there are several prevailing theories (see the
comprehensive reviews by M. Rees \cite{rees78} and M. Volonteri \cite{volonteri10a} for more details). One of the most 
popular possibilities is that the first black hole seeds are the remnants of the first generation of stars, the so-called 
population III stars, formed out of primordial ultra-low metallicity gas. These black holes formed at $z$$\sim$20 
and have typical masses $\sim$100-1,000~$M_\odot$. This scenario has problems explaining the very high masses, 
of $\sim$10$^9$M$_\odot$, estimated for supermassive black holes in $z$$\sim$6 optically-selected quasars \cite{willott10b}. 
Alternatively, the first black holes could have formed directly as the result of gas-dynamical processes. It 
is possible for metal-free gas clouds with $T_{\textnormal vir}$$\gsim$10$^4$K and suppressed $H_2$ formation to collapse 
very efficiently \cite{bromm03}, possibly forming massive black hole seeds with M$\sim$10$^4$-10$^5$M$_\odot$ as early 
as $z$$\sim$10-15. If instead the UV background is not enough to suppress the formation of $H_2$, the gas will fragment 
and form ``normal'' stars in a very compact star cluster. In that case, star collisions can lead to the formation of a very 
massive star, that will then collapse and form a massive black hole seed with mass $\sim$10$^2$-10$^4$M$_\odot$
\cite{devecchi09}.

Given the current masses of 10$^{6-9}$M$_\odot$, most black hole growth happens in the Active Galactic Nuclei (AGN) 
phase \cite{lynden-bell69,soltan82}. With typical bolometric luminosities $\sim$10$^{45-48}$erg~s$^{-1}$, AGN are amongst 
the most luminous emitters in the Universe, particularly at high energies and radio wavelengths. These luminosities are 
a significant fraction of the Eddington luminosity --- the maximum luminosity for spherical accretion beyond which radiation
pressure prevents further growth --- for a 10$^{8-9}$~M$_\odot$ central black hole. A significant fraction of the total black hole
growth, $\sim$60\% \cite{treister10}, happens in the most luminous AGN, quasars, which are likely triggered by the major merger 
of two massive galaxies \cite{sanders88}. In an AGN phase, which lasts $\sim$10$^8$ years, the central supermassive
black hole can gain up to $\sim$10$^7$-10$^8$ M$_\odot$, so even the most massive galaxies will have only a few of 
these events over their lifetime. Further black hole growth, mostly in low-luminosity (low Eddington rate) AGN, is 
likely due to stochastic accretion of cold gas, mostly in spiral galaxies \cite{hopkins06a}.

According to the AGN unification paradigm \cite{antonucci93,urry95}, a large fraction of these sources, $\sim$75\% locally, 
are heavily obscured by optically and geometrically thick axisymmetric material, which explains many of the observed 
differences among different types of active galaxies. In addition, luminosity \cite{lawrence91} and cosmic epoch
\cite{treister06b} play a significant role. One constraint on the fraction of obscured AGN and its evolution comes from the 
spectral shape of the extragalactic X-ray ``background'' (XRB). Thanks to deep X-ray observations at E$\lsim$10~keV 
performed by {\it Chandra} and {\it XMM-Newton}, a very large fraction of the X-ray background, $\sim$80\%, has been 
resolved into point sources \cite{hickox06}, the vast majority of them AGN \cite{mushotzky00}. Several studies, the first of 
them $\sim$20 years ago \cite{setti89}, have used a combination of obscured and unobscured AGN to explain the spectral 
shape and normalization of the X-ray background with overall good results. The latest AGN population synthesis 
models \cite{treister09b,draper09}  assume an average ratio of obscured to unobscured AGN of $\sim$3:1 locally, increasing
towards lower luminosities and higher redshifts, as well as a fraction of Compton-thick 
sources (CT; $N_H$$>$10$^{24}$cm$^{-2}$) of $\sim$5-10\%, consistent with the value observed at higher energies,
E=10-100~keV, of $\sim$5\% by INTEGRAL in the local Universe \cite{treister09b,burlon11}, lower by factors of $\sim$3
than expectations of previous population synthesis models \cite{treister05b,gilli07}. 

In this paper, we review multiwavelength methods used to trace the growth of SMBHs (\S2), the known properties of 
unobscured and obscured AGN (\S3 and \S4 respectively), the cosmic history of black hole accretion 
(\S5) and prospects for future observations (\S6). Throughout this paper, we assume a $\Lambda$CDM cosmology 
with $h_0$=0.7, $\Omega_m$=0.27 and $\Omega_\Lambda$=0.73, in agreement with the most recent 
cosmological observations \citep{hinshaw09}.

\section{How to trace SMBH Growth?}

\begin{figure}
\begin{center}
\includegraphics[scale=0.4]{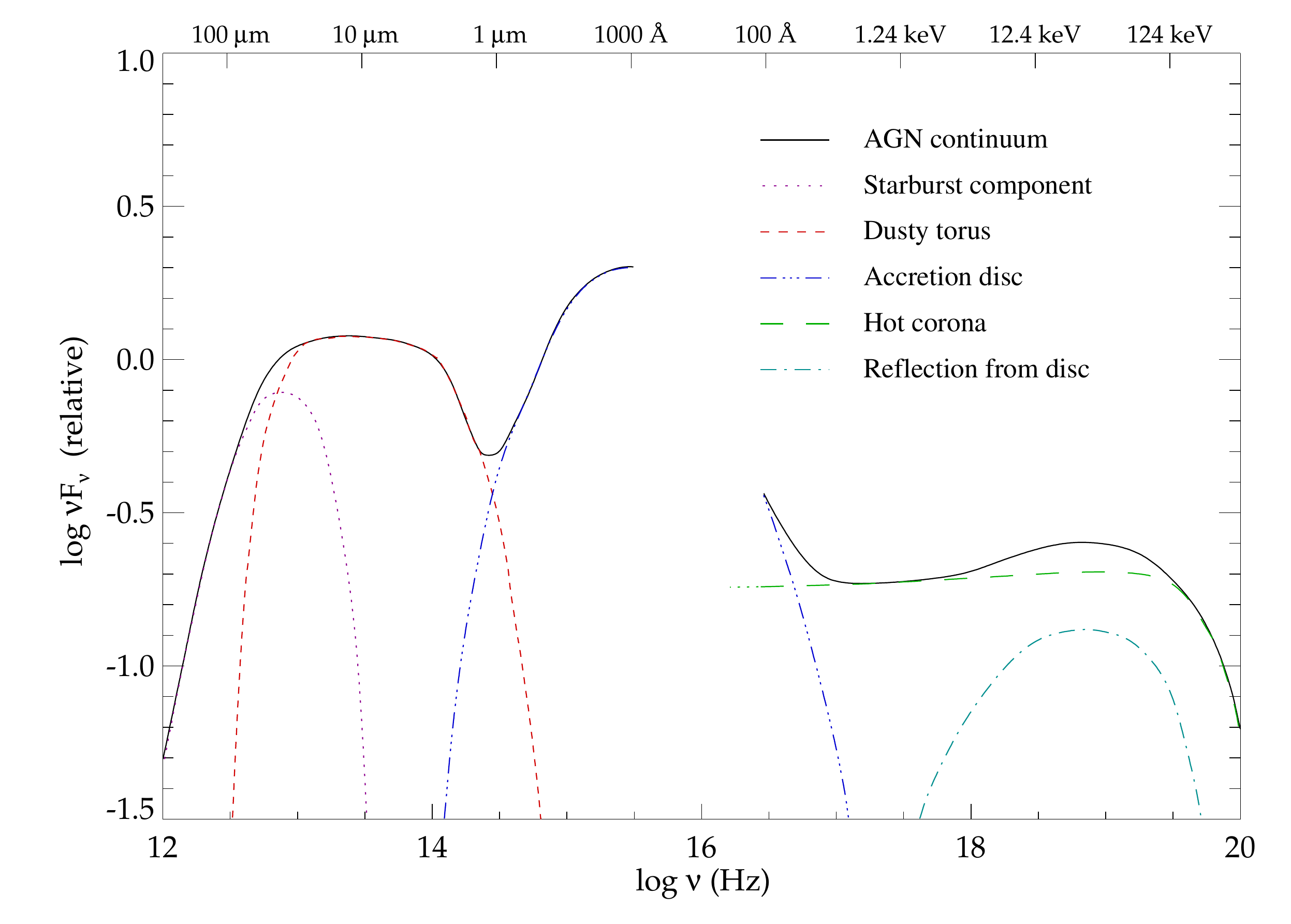}
\vspace{-0.7cm}
\caption{Average unobscured radio-quiet AGN spectrum from far infrared to hard X-rays, as compiled by Manners (2002; \cite{manners02}). The contributions 
from each component are shown separately together with the total AGN emission ({\it solid line}). Most of the AGN radiation 
appears in three regions: infrared (re-emission from dust; {\it dashed red line}), UV and optical (accretion disk; 
{\it dot-dot-dot-dashed blue line}), and X-rays (hot corona and reflection from the accretion disk; {\it
dashed green and dot-dashed cyan lines}). At longer wavelengths, $>$100~$\mu$m, a starburst component associated
with the host galaxy dominates ({\it dotted magenta line}).}
\label{agn_spec}
\end{center}
\end{figure}

One of the main features of the AGN emission is that it covers a very wide range of wavelengths, from radio to 
Gamma-rays (Fig.~\ref{agn_spec}). While unobscured sources are easily detectable and identified by their UV and soft X-ray
continuum and their broad optical emission lines, obscured AGN can only be found at longer, mid-IR, wavelengths or in 
hard X-rays. Of course, selections at different wavelengths have different biases. For example, while radio surveys are not 
particularly affected by obscuration, they are most likely to detect radio-loud sources, which are only $\sim$10\% of the 
total AGN population at bright fluxes \cite{condon98}. However, combining different multiwavelength techniques gets us 
closer to a complete picture. Below, we briefly describe the most popular AGN selection methods, their advantages 
and drawbacks.

\subsection{Optical/UV Continuum}

\begin{figure}
\begin{center}
\includegraphics[angle=270,scale=0.5]{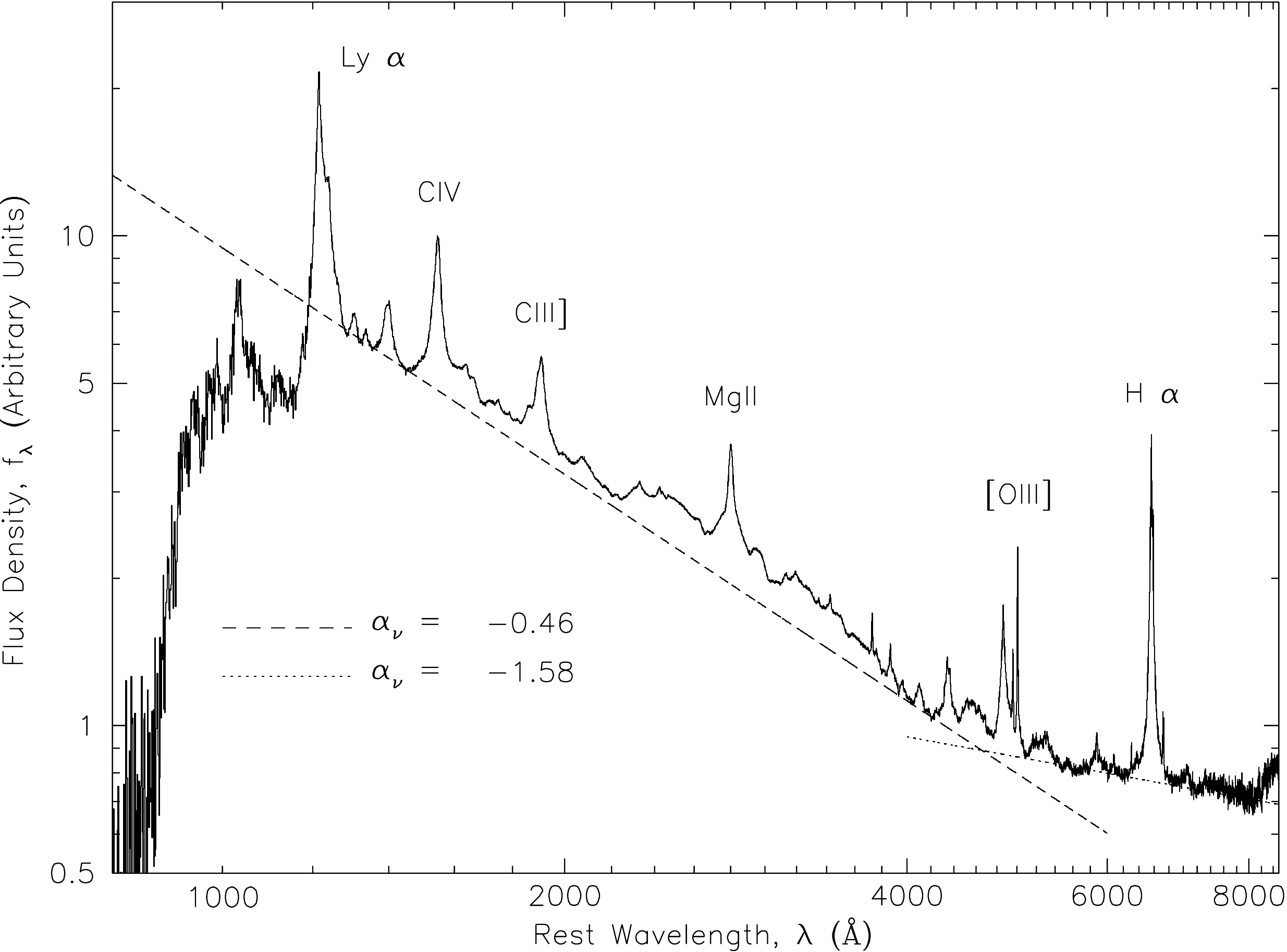}
\caption{Composite rest-frame optical/UV spectrum for the optically-selected quasars in the Sloan Digital Sky Survey, from the work
of Vanden Berk et al. \cite{vandenberk01}.
The {\it dashed} and {\it dotted} lines show power-law fits to the continuum emission.}
\label{sdss_spec}
\end{center}
\end{figure}

Rest-frame optical/UV selection of AGN, in particular for high-luminosity unobscured quasars, is particularly efficient 
because the spectral shapes of stars and quasars at these wavelengths produce very different broadband colors due to the
presence of the ``big blue bump'' \cite{shields78} in quasar spectra from $\sim$100\AA~to $\sim$1$\mu$m 
(Fig.~\ref{sdss_spec}). This emission is often attributed to the thermal radiation with temperatures $\sim$30,000K 
originating in the accretion disk \cite{malkan82}. This unique spectral shape has been used in the past to identify 
quasars with great success by optical surveys such as the Palomar-Green (PG) survey \cite{schmidt83}, the 
2 degree field QSO redshift survey \cite{boyle00} or more recently the Sloan Digital Sky Survey 
(SDSS; \cite{richards01}), which has found now more than 1 million quasars \cite{richards09}.

However, samples selected in the optical are far from complete, as emission at these wavelengths is strongly affected 
by reddening or extinction, and most AGN are obscured along our line-of-sight. Furthermore, for lower luminosity sources, 
the optical light from the host galaxy can easily outshine the nuclear emission. This is particularly important for 
ground-based observations and high-redshift sources, for which it is very hard to separate the non-thermal and
stellar components spatially.

\subsection{Radio}

Historically, identification of AGN based on their radio emission has been very important. In fact, the first discovered 
quasar, 3C 273, was originally classified as a radio source \cite{schmidt63}. In spite of this, radio selection can be very 
problematic. In radio-loud sources (defined as $f_{\textnormal 5GHz}$/$f_B$$>$10 \cite{kellermann89}) radio emission is 
associated with a strong, non-thermal, component, probably originating in a beamed collimated relativistic jet \cite{begelman84}. 
In radio-quiet sources, which are typically $\sim$3 orders of magnitude fainter at these wavelengths \cite{sanders89}, the
radio emission corresponds to the long-wavelength tail of the far-infrared dust emission. As a consequence, radio-selected 
samples are necessarily biased towards radio-loud sources, which represent only $\sim$10\% of the overall AGN
population.

\subsection{Optical Emission Lines}

\begin{figure}[h!]
\begin{center}
\includegraphics[angle=0,scale=0.8]{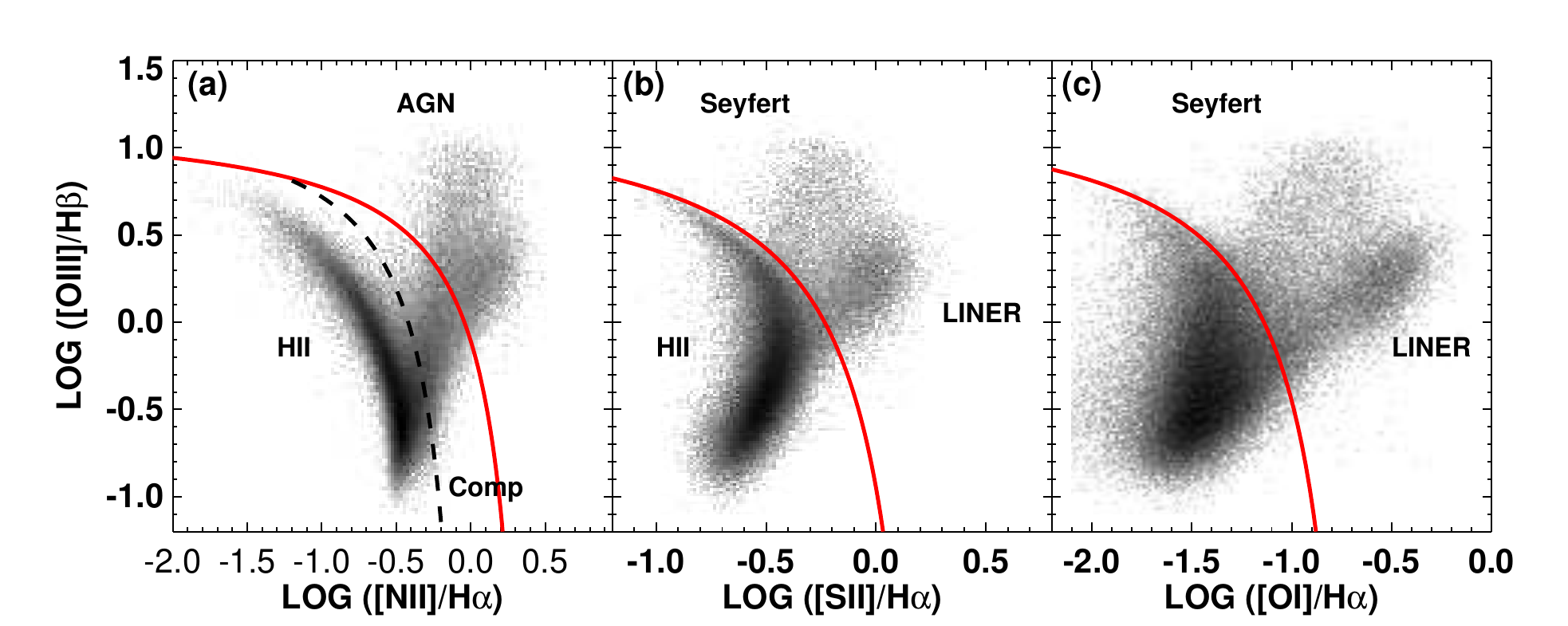}
\vspace{-0.5cm}
\caption{AGN selection diagrams based on optical emission line ratios. Figure obtained from Kewley et al. \cite{kewley06}. 
Divisions between star-forming galaxies and AGN are shown by the {\it dashed} \cite{kauffmann03} and 
{\it solid} \cite{kewley01} lines; bona-fide AGN sit at the upper right in these distributions.}
\label{bpt_diag}
\end{center}
\end{figure}

As first reported by Baldwin et al. \cite{baldwin81}, the photoionizing spectrum of a power-law continuum source, such as 
an AGN, produces very different emission line intensity ratios when compared with that of typical star-forming regions 
(mostly due to O and B stars). Hence, emission lines can be used to identify the presence of AGN even in galaxies in which 
the optical/continuum does not show any direct AGN signature, due to obscuration and/or low luminosity. Because the AGN 
ionizing emission reaches material even a few kilo-parsecs away from the nuclear region, this selection technique is less
sensitive to circumnuclear obscuration, and thus provides a more complete AGN view when compared with, for example, 
optical/UV continuum selection. This technique was used successfully in the SDSS \cite{kauffmann03,kewley06} to extend 
the low-redshift AGN sample to lower luminosities. Emission line ratios and diagnostic regions can be seen in 
Fig.~\ref{bpt_diag}. Emission-line selection can also be used at higher redshifts, as shown by the DEEP2 galaxy redshift 
survey, which selected a sample of 247 AGN at $z$$\sim$1 from optical spectroscopy using the DEIMOS spectrograph at 
the Keck observatory \cite{montero-dorta09}.

While this is an efficient AGN selection technique, optical spectroscopy is very expensive in telescope time, 
and is only feasible for relatively bright emission line regions. This selection may be incomplete at the low luminosity end, if 
the host galaxy can outshine the high-ionization emission lines. It is currently very difficult to extend this selection beyond 
$z$$\sim$1, as the relevant emission lines move to observed-frame near-IR wavelengths, where current-generation 
spectrographs are significantly affected by atmospheric emission, do not cover wide field of views and have limited 
multi-object capabilities.

\subsection{X-rays}

\begin{figure}
\begin{center}
\includegraphics[angle=270,scale=0.4]{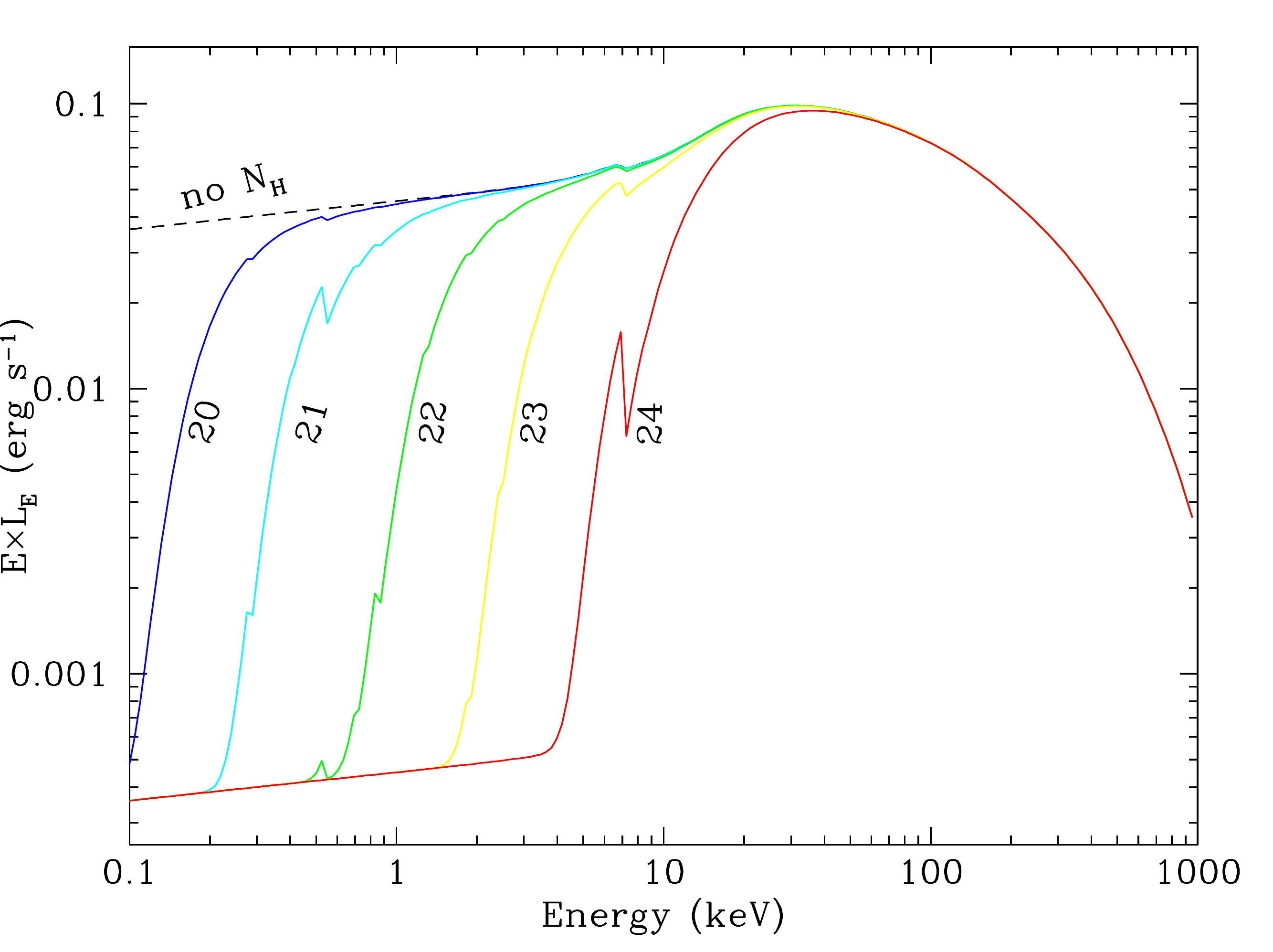}
\vspace{-0.4cm}
\caption{Typical AGN X-ray spectrum from 0.1 to 100 keV. Each spectrum includes a power-law with photon index 
$\Gamma$=1.9, a Compton reflection hump, peaking at $\sim$30 keV, a high-energy cutoff at 300 keV, and photoelectric 
absorption. Numbers for each curve indicate the amount of absorption, given as $\log$(N$_H$), with $N_H$ in units of 
atoms~cm$^{-2}$.}
\label{xspec}
\end{center}
\end{figure}

As was found more than 30 years ago, AGN are ubiquitous X-ray emitters \cite{elvis78}. Their X-ray emission extends from
$\sim$0.1 keV to $\sim$300 keV and is attributed to inverse-Compton scattering due to high-energy electrons in a hot
corona, surrounding the accretion disk. The high-energy cutoff at $\sim$100-300 keV is presumably due to a cutoff in the 
energy distribution of the electrons in the hot corona. AGN are typically $\sim$1-5 orders of magnitude more luminous in 
X-rays than normal galaxies, which makes them the dominant extragalactic population at these wavelengths. Most AGN are
obscured by photoelectric absorption by gas along the line of sight, which preferentially affects emission at the lower 
energies. This is often parametrized by the amount of neutral hydrogen column density in the line of sight. Fig.~\ref{xspec} 
shows typical AGN X-ray spectra including the power-law component and high energy cutoff, for different levels of 
photoelectric absorption.

Deep X-ray surveys with {\it Chandra} and {\it XMM-Newton} have found the largest AGN densities, 
$\sim$7,000 deg$^{-2}$ \cite{bauer04}, $\sim$10-20 times higher than even the deepest optical surveys. Still, X-ray 
selected AGN samples are still biased against the most obscured sources. In fact, even the deepest {\it Chandra} surveys 
can miss more than half of the total AGN due to a combination of obscuration and/or low luminosity \cite{treister04}. Observing 
at higher energies helps to alleviate the effects of obscuration. Wide-area surveys performed by the {\it Swift} \cite{tueller08} 
and {\it INTEGRAL} \cite{sazonov07} satellites have detected a large number of AGN in the local Universe at 
$E$=10-100~keV, where all but the most obscured, Compton-thick, sources emit strongly. Unfortunately, due to their 
relatively high flux limits ($\sim$2-3 order of magnitudes shallower than the deeper {\it Chandra} observations), surveys at 
these energies are so far limited to low redshifts only.

\subsection{Infrared}

Intrinsic AGN emission is not particularly strong at near/mid-IR wavelengths. Radiation coming from the accretion disk, 
often characterized as a power-law, while very strong at UV and optical wavelengths decreases rapidly 
beyond $\sim$1~$\mu$m \cite{richards06}. However, as was originally found by {\it IRAS} \cite{sanders89} and latter 
confirmed by {\it ISO} \cite{haas03} and {\it Spitzer} \cite{treister06a}, AGN are luminous IR sources. This is commonly 
attributed to re-emission of absorbed energy by dust. This radiation can be found starting at $\sim$2-3~$\mu$m, which
corresponds to the dust sublimation temperature, about $\sim$1,000-2,000~K \cite{glikman06}, and extends to 
$\sim$100~$\mu$m, at the tail of the black body spectrum for a $\sim$100-1,000 K distribution \cite{nenkova02}, where
the emission from the host galaxy typically starts to dominate. Typical AGN infrared luminosities are 
10$^{44-46}$~erg~s$^{-1}$ and thus they represent a significant fraction, $\sim$30\% on average \cite{treister08}, of 
the bolometric luminosity.

One clear advantage of infrared AGN selection is that this re-emission is mostly isotropic, and hence both obscured and 
unobscured sources have similar detection probabilities. Therefore, it provides a complementary approach to the most 
common selection techniques described above, which are less sensitive to obscured sources. However, star-forming galaxies 
are very luminous at these wavelengths as well, so that the host galaxy can easily outshine the central emission, in particular 
for low-luminosity sources, hence yielding a low efficiency in detecting AGN \cite{cardamone08}. Furthermore, the selection 
function in infrared studies is more complicated, as the probability of detecting the AGN depends on the properties of the host 
galaxy, such as the amount of dusty star-formation, which is in principle independent of the nuclear luminosity.

\section{Unobscured AGN}
\label{unobscured}

Because quasars are the most luminous, and thus easily detectable, members of the AGN family, the luminosity function of
optical quasars has been well determined for years. In particular, it was found that the number of quasars evolve 
strongly \cite{schmidt68} and peak at $z$$\simeq$2 \cite{marshall85}. This evolution has been modeled either as a pure 
luminosity evolution (PLE), in which the characteristic luminosity changes with redshift while the shape of the luminosity 
function remains the same \cite{mathez76}, or a pure density evolution (PDE), so that only the normalization of the luminosity
function depends on redshift \cite{schmidt68}; however, it was quickly discovered that at least in the PG quasar survey, the 
shape of luminosity function also evolves with redshift and thus neither a PLE nor a PDE provides a good description 
\cite{schmidt83}.

\begin{figure}
\begin{center}
\includegraphics[angle=270,scale=0.4]{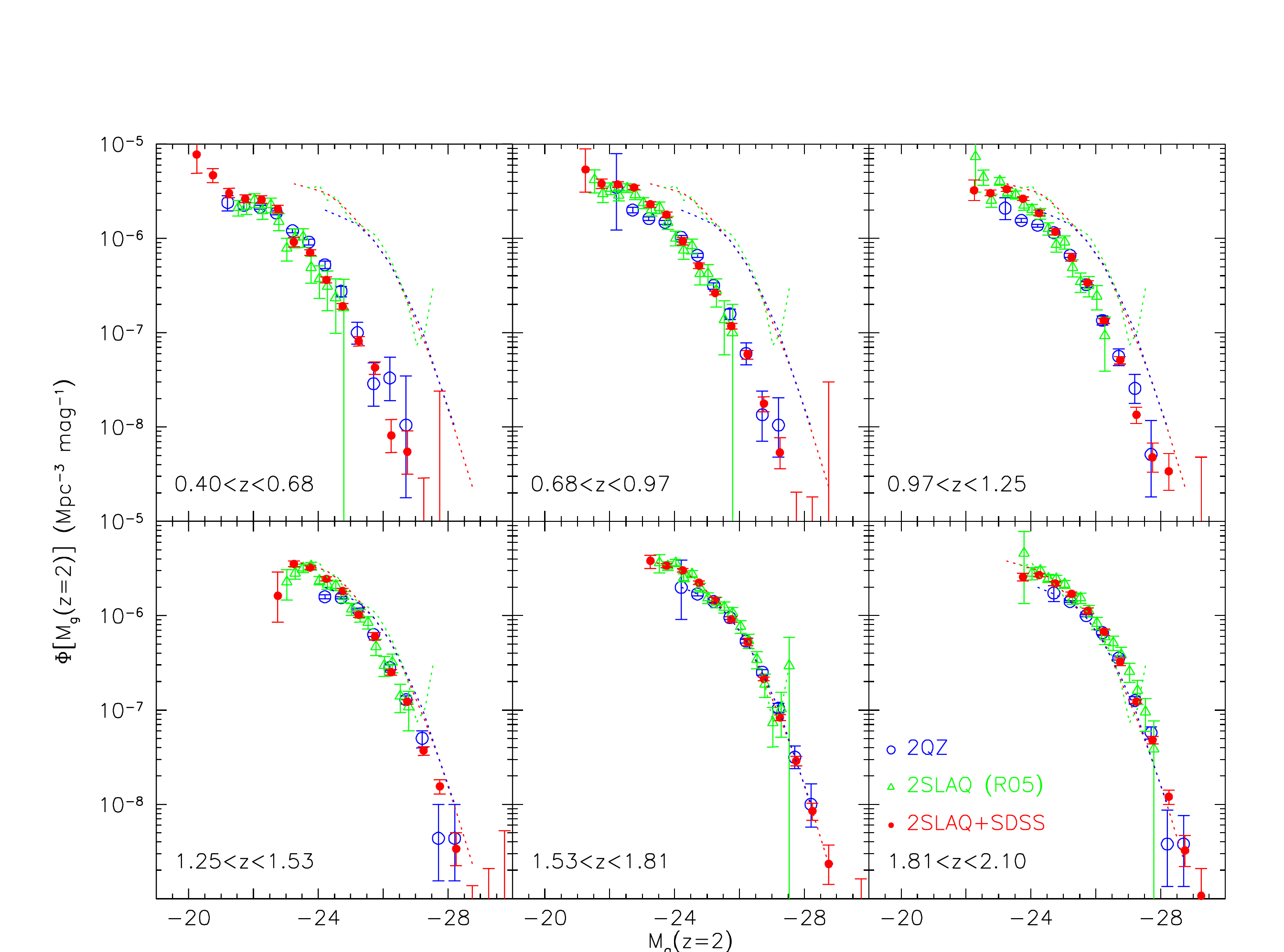}
\vspace{-0.3cm}
\caption{Binned optical quasar luminosity function from the work of Croom et al. (\cite{croom09}; their Fig.~11) in six
redshift intervals, from $z$=0.4 to $z$=2.1. Measurements where obtained from the 2dF QSO Redshift Survey 
(2QZ; \cite{croom04} {\it blue}), the 2dF-SDSS LRG and QSO survey (2SLAQ; \cite{richards05}; {\it green}) and combining
the latter with the general SDSS QSO sample (\cite{richards06b}; {\it red}).The dotted line shows the measured luminosity
function in the 1.53$<$$z$$<$1.81 range for reference. The strong evolution of the QSO luminosity function is clearly seen
in this figure. This evolution is best fitted by a LEDE model, as described in the text.}
\label{qso_lf_croom09}
\end{center}
\end{figure}

In Fig.~\ref{qso_lf_croom09} we show one of the latest measurements of the quasar luminosity function
at 0.4$<$$z$$<$2.6, from Croom et al. \cite{croom09}. They conclude that a luminosity-dependent density evolution provides 
a better fit to the 
optical quasar luminosity function, than either a PLE or PDE. Similar conclusions were reached by studying X-ray selected 
sources using soft X-ray (0.5-2 keV) observations \cite{hasinger05} or hard X-ray (2-10 keV) data \cite{ueda03,yencho09} 
which includes both obscured and unobscured AGN. However, taking advantage of the large number of sources in their 
sample, $\sim$10,000, Croom et al. \cite{croom09} concluded that the best fit to the observed luminosity function is obtained 
by using a model based on a luminosity evolution + density evolution (LEDE). The most important difference between
a LEDE and a PLE fit is a change in amplitude and bright-end slope at high redshift.

\begin{figure}
\begin{center}
\includegraphics[angle=0,scale=2.5]{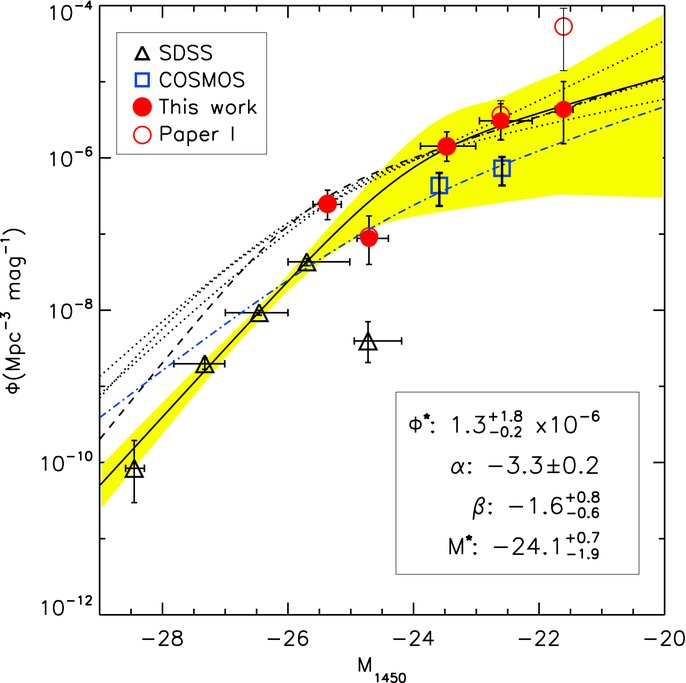}
\includegraphics[angle=0,scale=0.4]{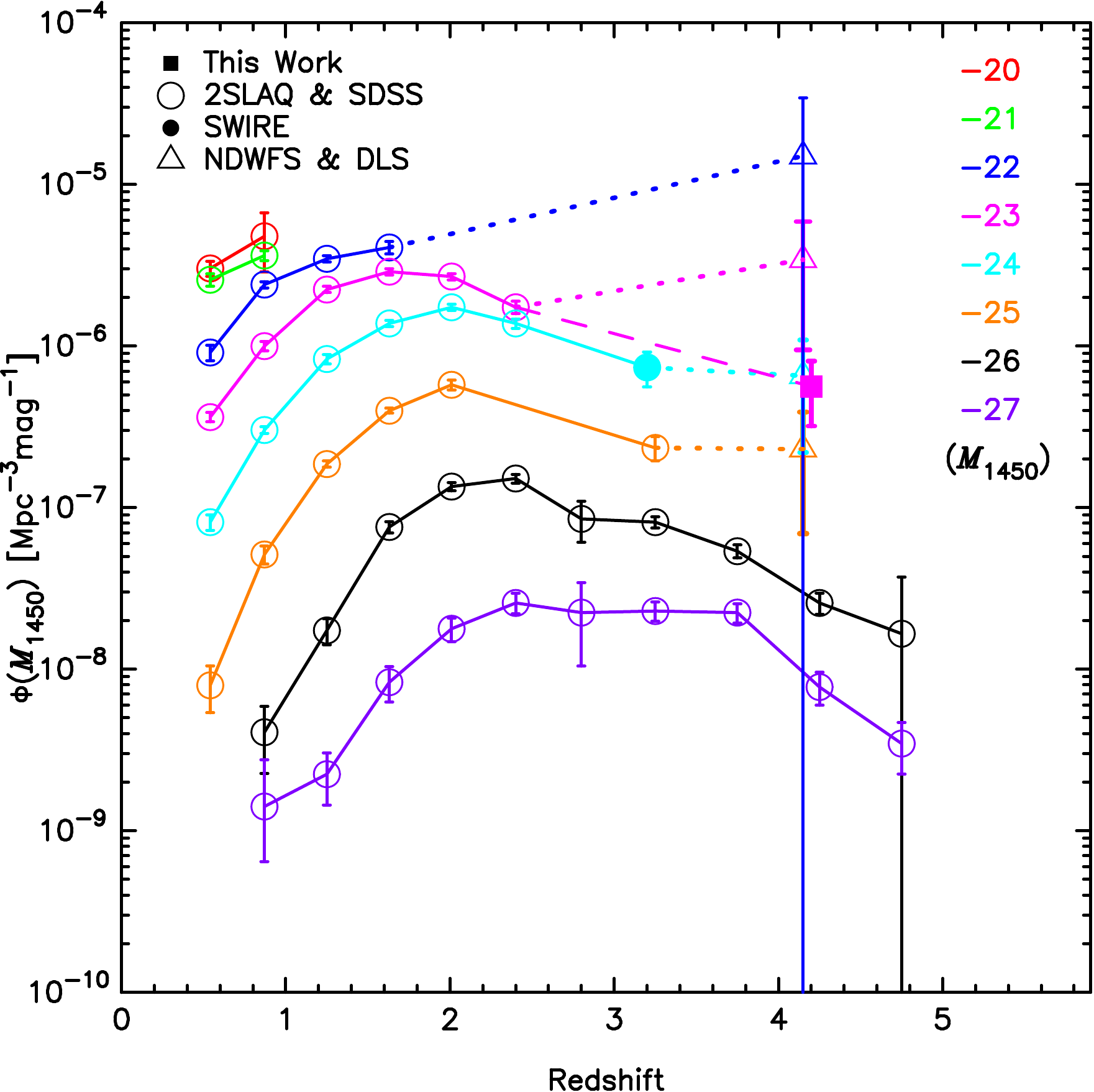}
\vspace{-0.4cm}
\caption{{\it Left panel:} $z$$\sim$4 quasar luminosity function, from the work of Glikman et al. \cite{glikman11}. The
{\it filled red circles} show the measurements presented on that work, while the {\it open red circles} were reported previously 
by the same group \cite{glikman10} and show the change with increased spectroscopic completeness. The {\it triangles} were 
obtained from the SDSS quasar sample at $z$=4.25 \cite{richards06b}. The {\it blue squares} are the space densities of 
$z$$\sim$4 QSOs from Ikeda et al. \cite{ikeda11} and the {\it dot-dashed line} is their best-fit double power law. The lower 
right-hand legend lists the best-fit parameters to a double power-law ({\it solid line}) shaded region represents the 1$\sigma$ 
uncertainties. Dashed and dotted lines show the $z$$\sim$3 QLF \cite{siana08}, representing different fits to the observed QLF.
{\it Right panel}: Quasar space density as a function of redshift from the work of Ikeda et al. \cite{ikeda11}. {\it Dotted lines} used 
the combined 2SLAQ, SDSS, SWIRE, NDFWS and DLS samples, while the {\it dashed lines} combine the COMOS and 2SLAQ 
sources. While AGN downsizing is clearly visible at $z$$<$2.5, at higher redshifts the situation is more uncertain.}
\label{lf_ikeda11}
\end{center}
\end{figure}

An important conclusion obtained from observations of the quasar luminosity function is the evidence for ``cosmic downsizing''
\cite{barger05}, i.e., that the most massive black holes get most of their mass at high redshift, while at low redshift only low 
mass black holes are still growing. This is observed both in optical \cite{croom09} and hard X-ray luminosity 
functions \cite{ueda03,barger05}, thus indicating that this result is independent of obscuration. Recent deep optical surveys 
such as the Great Observatories Origins Deep Survey (GOODS; \cite{cristiani04}), the Cosmic Evolution Survey 
(COSMOS; \cite{ikeda11}), the NOAO Deep Wide Field Survey (NDWFS) and the Deep Lens Survey (DLS; \cite{glikman10})
have produced significant advances in extending the quasar luminosity function to higher redshifts, $z$$>$3. 
Fig.~\ref{lf_ikeda11} shows the quasar luminosity function at $z$$\sim$4 and the redshift dependence of the quasar spatial
density \cite{cristiani04,glikman10}. While the presence of downsizing is clear up to $z$$\sim$2.5, at higher redshifts it is less 
convincing, most likely due to poor statistics and incompleteness. As argued by Glikman et al. \cite{glikman11}, the slope of 
the faint end of the quasar luminosity function is critical in determining the contribution of these sources to the ionization of the 
intergalactic medium. Based on current measurements, quasars contribute $\sim$60\% of the ionizing photons at $z$$\sim$4 
and thus are the dominant source at this redshift.

At even higher redshifts, $z$$\sim$5-6, current deep surveys do not cover enough area to detect a significant number of sources.
However, wide-area survey such as SDSS \cite{fan01a} and the Canada-France High-z Quasar Survey \cite{willott10a} have 
been able to find a sizable sample, $\sim$40 high-luminosity quasars, at these high redshifts. According to these samples, 
there is a large decrease in the number density of high-redshift quasars, when compared to $z$$\sim$2, suggesting that the 
peak of the quasar activity is at $z$$\sim$2.5 \cite{richards06b}. The decline towards higher redshifts is given by 
10$^{-0.47z}$ \cite{fan01a} from $z$=3 to $z$=6. A similar trend is also observed for high luminosity sources in X-ray 
surveys \cite{civano11}, which are also dominated by unobscured sources. This indicates that due to their very low spatial 
density, unobscured quasars do not contribute significantly to the early hydrogen re-ionization of the intergalactic medium 
at $z$$\sim$6 \cite{fan01a,willott10a}, in contrast to the situation at $z$$\sim$4.

\section{Obscured Accretion}
\label{obscured}

\begin{figure}
\begin{center}
\includegraphics[angle=0,scale=0.4]{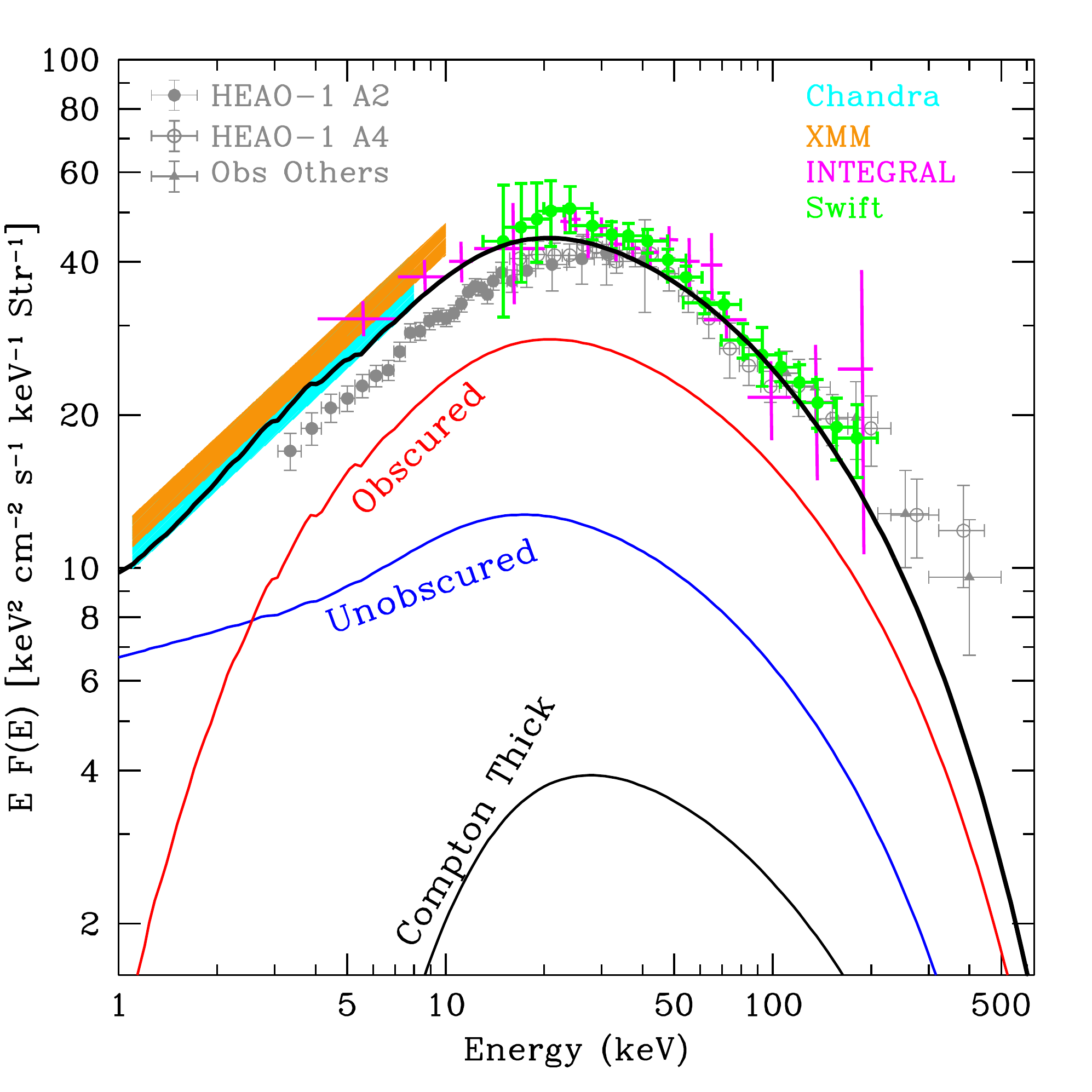}
\vspace{-0.6cm}
\caption{Observed spectrum of the extragalactic XRB from HEAO-1 \cite{gruber99}, Chandra \cite{hickox06}, 
XMM \cite{deluca04}, INTEGRAL \cite{churazov07} and Swift \cite{ajello08} data. The {\it thick black solid line} shows the 
population synthesis model for the XRB spectrum of Treister et al. \cite{treister09b}. {\it Red}, {\it blue} and {\it thin black} 
solid lines show the contribution to this model from unobscured, obscured Compton thin and CT AGN respectively.}
\label{xrb}
\end{center}
\end{figure}

The space density and evolution of the unobscured AGN population has been well studied, mostly from optical
and soft X-ray surveys. However, we know that a large fraction of the SMBH growth happens in heavily obscured systems.
Observations of the nearest AGN suggest that the local ratio of obscured to unobscured sources is $\sim$4:1 \cite{maiolino95}.
A similarly high fraction of obscured AGN has been used to explain the spectrum and normalization of the extragalactic XRB, as 
shown by the latest AGN population synthesis models \cite{treister05b,ballantyne06,gilli07,treister09b}. The XRB gives an integral 
constraint to the AGN population and its evolution; the most recent deep surveys show that $\sim$90\% of the observed 2--8
keV XRB radiation can be attributed to resolved AGN \cite{hickox06}. In Fig.~\ref{xrb}, we show the latest AGN population 
synthesis models for the XRB which uses a local ratio of obscured to unosbcured AGN of $\sim$3:1, plus a luminosity and 
redshift dependence, as described below \cite{treister09b}. The largest uncertainty in such models stems from the
normalization mismatch of the data in the 1-10 keV energy range.

\begin{figure}
\begin{center}
\includegraphics[angle=0,scale=0.3]{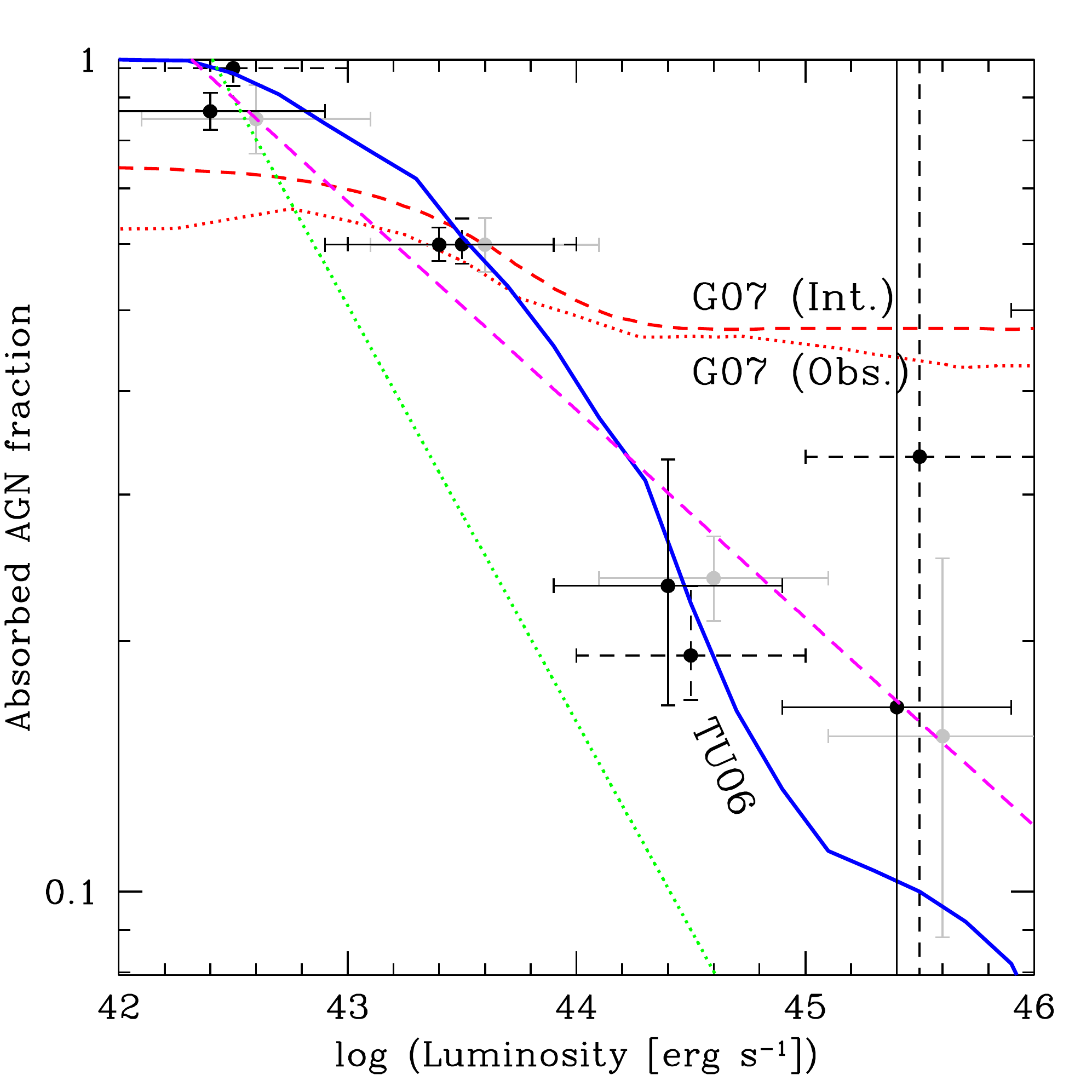}
\includegraphics[angle=0,scale=0.56]{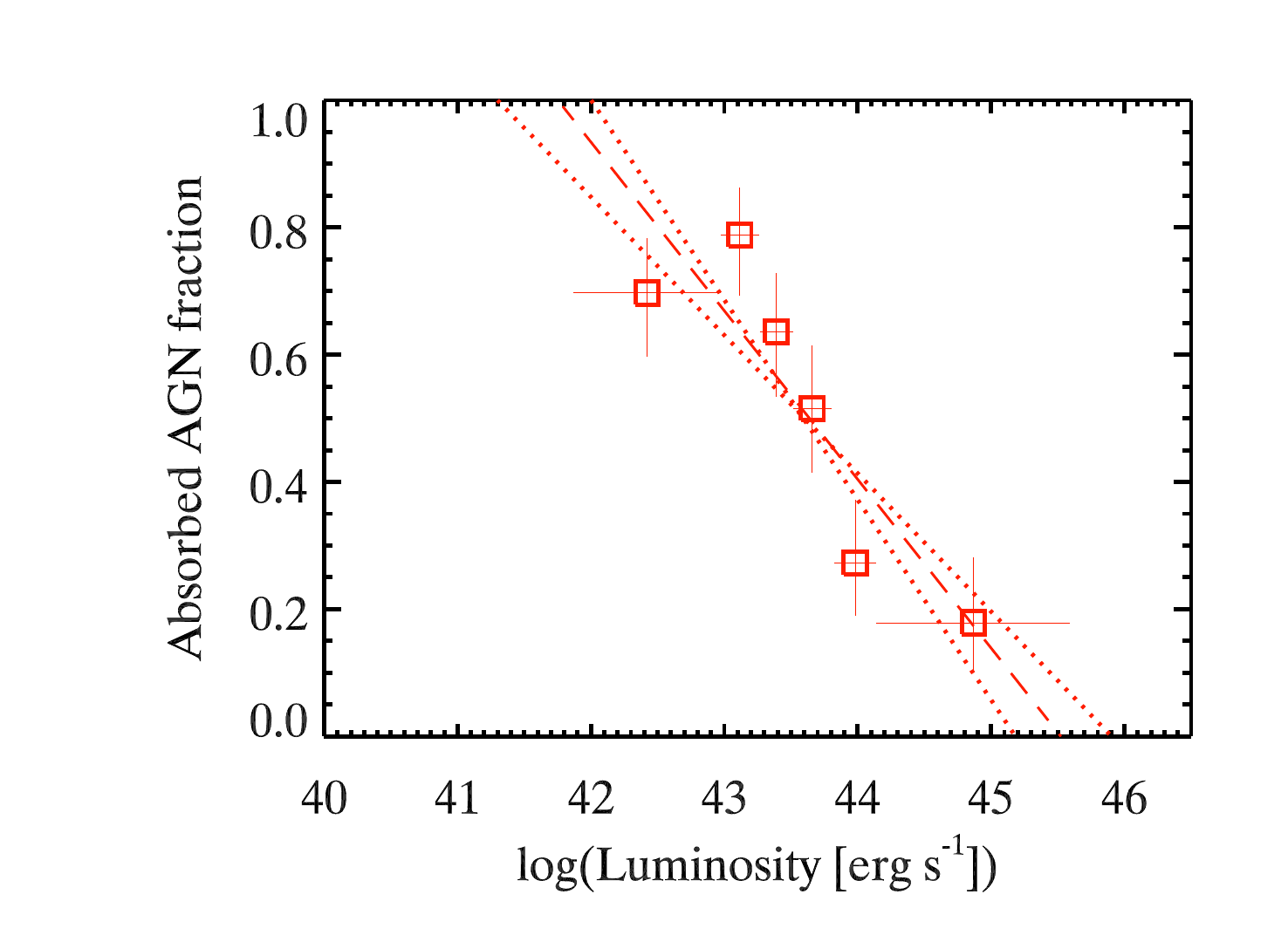}
\vspace{-0.5cm}
\caption{{\it Left panel}: Ratio of obscured to total AGN as a function of hard X-ray luminosity. The black circles with 
{\it dashed error bars} show the obscured AGN fraction from the extended {\it Chandra} deep field south (ECDF-S) alone, 
while the {\it gray circles} show the results obtained using the meta analysis by Treister \& Urry \cite{treister06b}. Black 
circles with {\it solid error bars} show the fraction obtained combining these two samples. The dependence in the AGN 
population synthesis model of Gilli et al. \cite{gilli07} for the intrinsic and observed fractions of obscured AGN are shown 
by the {\it dashed} and {\it dotted} red lines; the dependence used by Treister \& Urry \cite{treister06b} is shown by the 
{\it solid blue line}. The {\it dashed magenta line} shows the expected dependence for a radiation-limited torus \cite{hoenig07}, 
while the {\it dotted green line} shows the expectation for the original ``receding torus'' \cite{lawrence91}, both normalized to the 
observed value in the 10$^{42-43}$~erg~s$^{-1}$ bin. {\it Right panel}: Same as left panel but using an AGN sample 
at $z$$\sim$0 selected in hard X-rays from {\it Swift}/BAT observations \cite{burlon11}. The fact that the same
luminosity dependence is observed in both samples indicates that it is not due to selection effects.}
\label{lum_ratio}
\end{center}
\end{figure}

A possible dependence of the fraction of obscured AGN on luminosity was first suggested nearly 20 years 
ago \cite{lawrence91}, and confirmed since by hard X-ray surveys \cite{ueda03,steffen03,barger05}. A possible 
physical explanation, is the so-called ``receding torus,'' in which the size of the inner opening angle depends on luminosity 
\cite{lawrence91,simpson05}. More recent observations found a luminosity dependence of the ratio of mid-IR to bolometric flux 
for unobscured AGN, consistent with this idea \cite{treister08}. Alternatively, it has been proposed that the observed 
dependence of the obscured fraction on luminosity could be explained either by the effects of photo-ionization on the X-ray 
obscuring matter \cite{akylas08} or by the Eddington limit on a clumpy torus \cite{hoenig07}. In Fig.~\ref{lum_ratio} we show 
the observed fraction of obscured AGN as a function of luminosity obtained by combining data from deep {\it Chandra} X-ray
surveys \cite{treister09a}. A consistent result is observed for AGN selected in a hard X-ray ($E$=14-195 keV) survey,
as shown in the right panel of Fig.~\ref{lum_ratio}, indicating that this trend is not due to selection biases.

In the left panel of Fig.~\ref{lum_ratio} we compare the observed dependence of the fraction of obscured sources on luminosity
with the expectations for different geometrical parameters of the obscuring material. If the height of the torus is roughly 
independent of luminosity, the change in covering fraction is due to a change in inner radius (the original ``receding torus'' 
model), hence a rough $L^{-1/2}$ dependence for the contrast should be expected \cite{barvainis87,lawrence91}. If the effects 
of radiation pressure are incorporated, in the case of a clumpy torus, a $L^{-1/4}$ dependence is expected. As can be seen 
in Fig.~\ref{lum_ratio}, a $L^{-1/2}$ dependence is too steep compared to observed data. This implies that the height of the 
obscuring material cannot be independent of the source luminosity and provides evidence for a radiation-limited structure.

\begin{figure}
\begin{center}
\includegraphics[angle=0,scale=0.4]{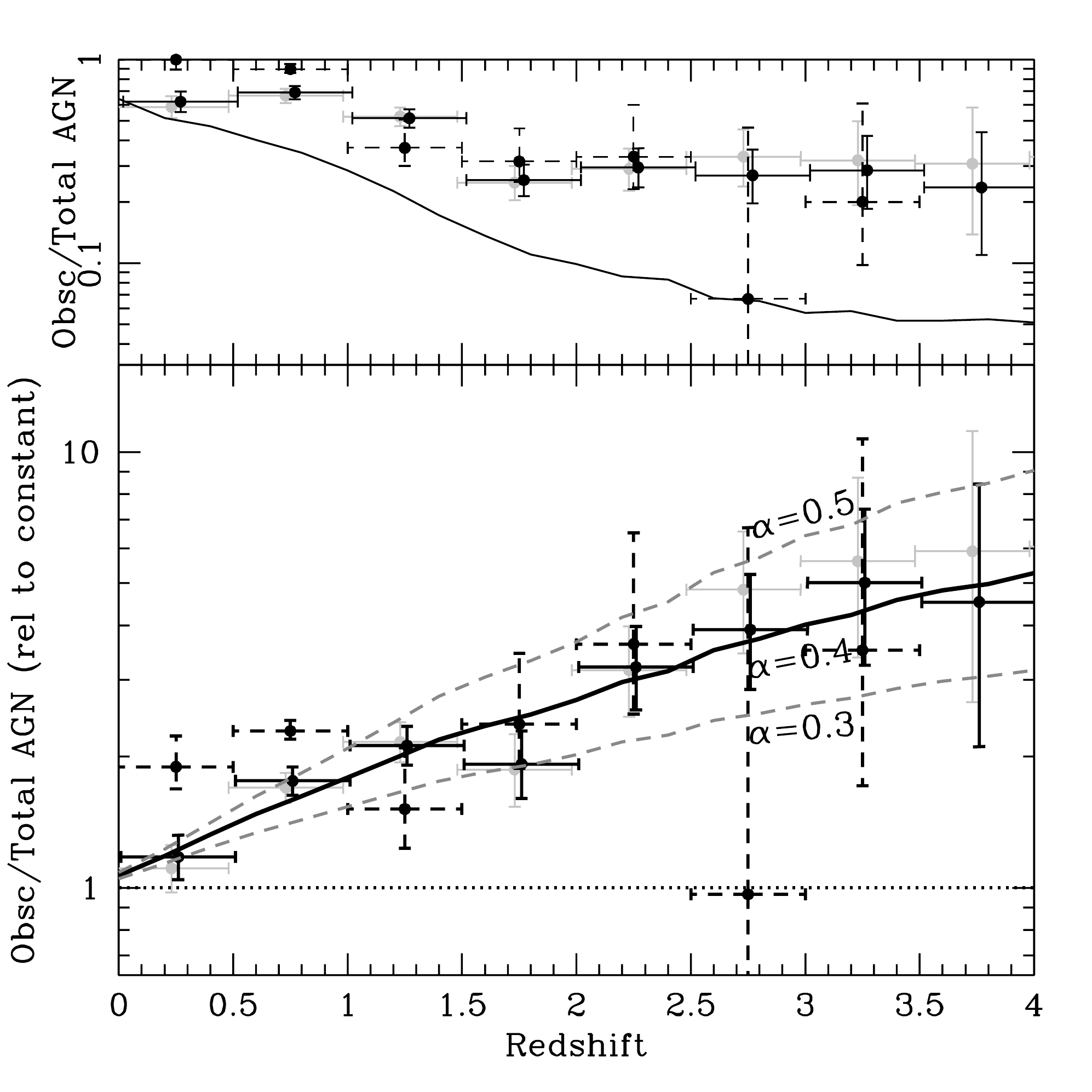}
\vspace{-0.7cm}
\caption{Fraction of obscured AGN as a function of redshift. {\it Upper panel}: direct measurements 
using the sources on the ECDF-S field only ({\it black circles with dashed error bars}),
from the sample of Treister \& Urry (\cite{treister06b}; {\it gray circles}) and combining both samples ({\it
black circles with solid error bars}). The expected observed fraction for an intrinsic
fraction of 3:1 obscured to unobscured AGN, accounting for optical and X-ray selection effects,
is shown by the {\it black solid line}. As can be seen, while the observed fraction of
obscured AGN declines toward higher redshifts, if the X-ray and optical selection effects
and the luminosity dependence of the obscured AGN fraction are taken into account, this
decline should be even stronger. {\it Bottom panel}: Inferred fraction of obscured AGN
relative to an intrinsically constant fraction after correcting for selection effect and
including the luminosity dependence of the obscured AGN fraction. Symbols are the same as
for the upper panel. The corrected fraction of obscured AGN increases with redshift
following a power-law of the form (1+$z$)$^\alpha$ with $\alpha$=0.4$\pm$0.1.}
\label{red_ratio}
\end{center}
\end{figure}

The dependence of the fraction of obscured AGN on redshift is more controversial. While some studies 
\cite{lafranca05,ballantyne06,treister06b,dellaceca08a,hasinger08} found a small increase in the fraction of obscured AGN at
higher redshifts, other results suggest that this fraction is constant \cite{ueda03,akylas06}. These discrepancies can
be understood due to a combination of small samples and the use of $N_H$ to classify AGN, which produces a well-know
redshift bias \cite{akylas06}. The fraction of obscured AGN as a function of redshift for a large, $\sim$2,000 sources, X-ray 
selected sample \cite{treister09a}, using optical emission lines to separate obscured and unobscured AGN, is shown in 
Fig.~\ref{red_ratio}. It increases significantly with redshift, roughly as (1+$z$)$^\alpha$, with $\alpha$=0.3-0.5 (thin dashed lines, 
bottom panel, Fig.~\ref{red_ratio}; best fit, $\alpha$$\simeq$0.4, thick dashed line).  This value of $\alpha$ does not change 
significantly if a different host galaxy evolution is assumed, and it is consistent with the value of 0.3 reported by other 
studies \cite{lafranca05,ballantyne06,hasinger08}.

Since star-forming galaxies may be expected to have more dust, the increase in the relative fraction of obscured AGN at high 
redshift may be due to an increase in the contribution to obscuration by galactic dust. By combining hard X-ray and mid-infrared 
observations, a similar ratio of hard X-ray to mid-infrared flux for obscured and unobscured AGN has been found \cite{lutz04}, 
contrary to the predictions of the simplest AGN unification paradigm, in which the obscuration comes from the dusty torus and 
therefore the mid-infrared emission is reduced due to self-absorption. This result can be explained if the obscuration comes 
from a much more extended region, i.e., kiloparsec, galactic scales rather than a compact parsec-scale torus. Furthermore, 
signatures for extended absorbing regions have been detected in nearby galaxies like NGC 1068 \cite{bock98} and NGC 4151 
\cite{radomski03}. Heavy absorption at kiloparsec-scales has routinely been found in ultra-luminous infrared galaxies 
(ULIRGs), which suffer a very strong evolution \cite{saunders90}. Hence, it seems likely that the change in the relative 
fraction of obscured AGN could be related to galactic-scale absorption, in particular since some ULIRGs also contain an 
obscured AGN (e.g., Arp 220; \cite{iwasawa05}).

Below, we review in more detail our current knowledge of the obscured AGN population at three different cosmic 
epochs: $z$$\simeq$0, $z$=1-3 and $z$$>$6.

\subsection{Obscured AGN in the Local Universe}

Nearby AGN are found in the so-called Seyfert galaxies \cite{seyfert43}, which are know to host low luminosity and/or obscured
nuclear activity \cite{maiolino95,risaliti99}. These growing supermassive black holes have been identified because of their 
high-ionization optical emission lines and in some cases their blue UV/optical continuum. The first nearby AGN catalogues, 
produced $\sim$40 years ago \cite{de-veny71}, contained $\sim$200 quasars. Roughly speaking, $\sim$5-15\% of the galaxies 
near the Milky Way contain an active nucleus \cite{maiolino95}, and $\sim$75\% of these active galaxies are obscured. In fact,
two of the three nearest AGN are Compton-thick (NGC 4945 and the Circinus galaxy \cite{matt00}). Hence, optical surveys are
not particularly efficient in unveiling this accretion, while observations at other wavelengths, in particular in the 
infrared \cite{de-grijp85}, and hard X-rays, are more complete.

Surveys at hard X-ray energies, E$>$15~keV, have been very successful in providing the most complete AGN samples 
in the local Universe. As long at the neutral hydrogen column density is lower than $\sim$10$^{24}$~cm$^{-2}$, the direct AGN
emission is mostly unaffected at these energies. Current observations at E$>$10~keV with the International Gamma-Ray 
Astrophysics Laboratory (INTEGRAL; \cite{winkler03}) and Swift \cite{gehrels04} satellites are available only at relatively high 
fluxes, and hence low redshifts, $z$$<$0.05. 

Using the IBIS coded-mask telescope \cite{ubertini03}, INTEGRAL surveyed $\sim$80\% of the sky down to a flux of 5 
mCrab in the 17-60~keV. Krivonos et al. \cite{krivonos07} report the properties of 130 AGN detected in these all-sky 
observations. A large number of unidentified sources remain in this full INTEGRAL catalog (48) but only seven are 
found at high galactic latitude ($|$$b$$|$$>$5$^o$), and thus are likely of extragalactic origin. Five of the 130 known AGN 
are Compton thick. Using similar observations from the all-sky Swift/BAT survey, a catalog of 103 AGN \cite{tueller08} contains 
five AGN with estimated $N_H$ greater than 10$^{24}$~cm$^{-2}$. However, we caution that some of these $N_H$ 
measurements were obtained by fitting a single absorbed power-law to the X-ray spectrum, while heavily absorbed AGN 
have more complex spectra \cite{vignati99,levenson06}, so the $N_H$ estimates are likely to be lower limits.

\begin{figure}
\begin{center}
\includegraphics[angle=270,scale=0.31]{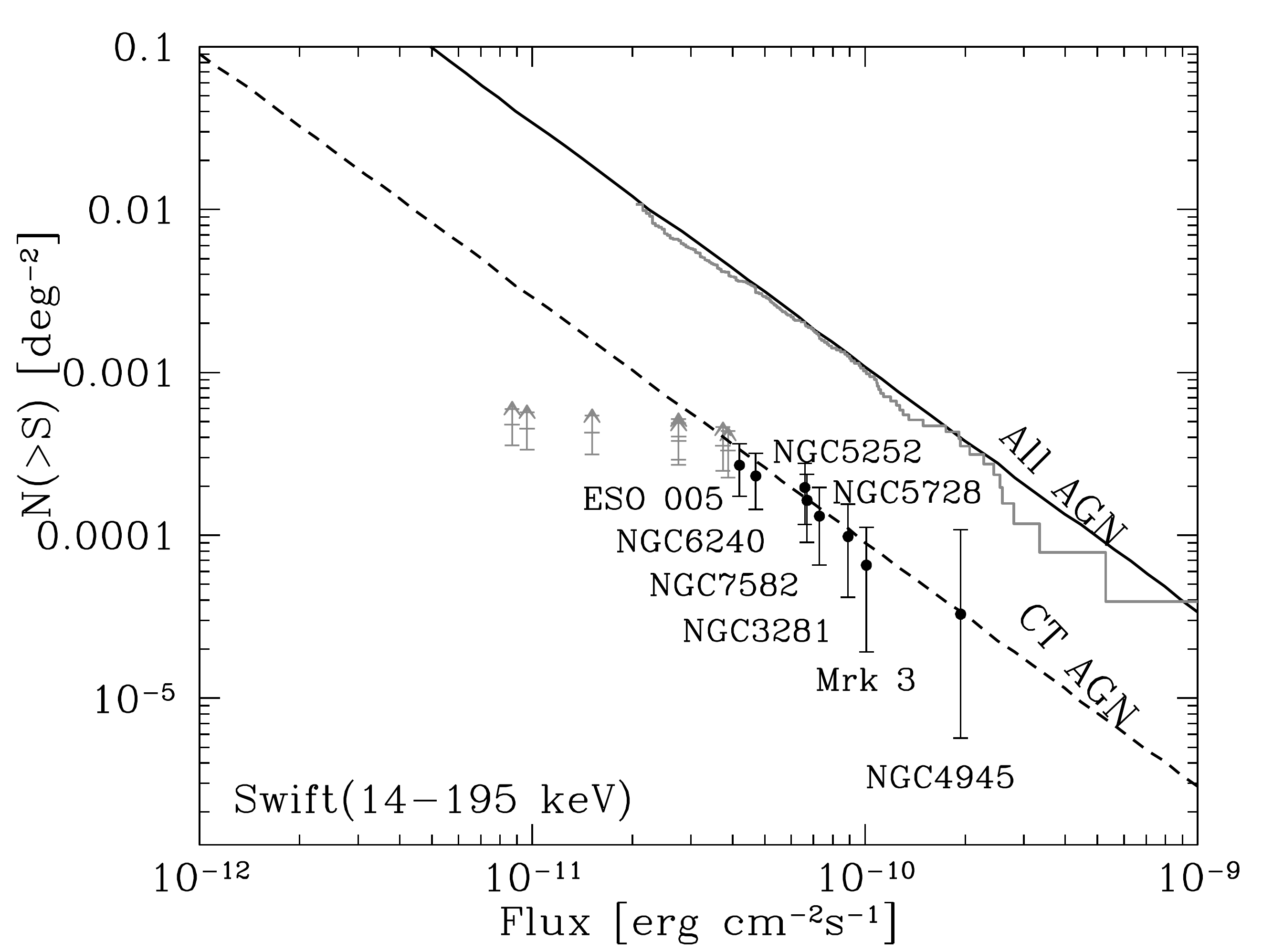}
\includegraphics[angle=270,scale=0.31]{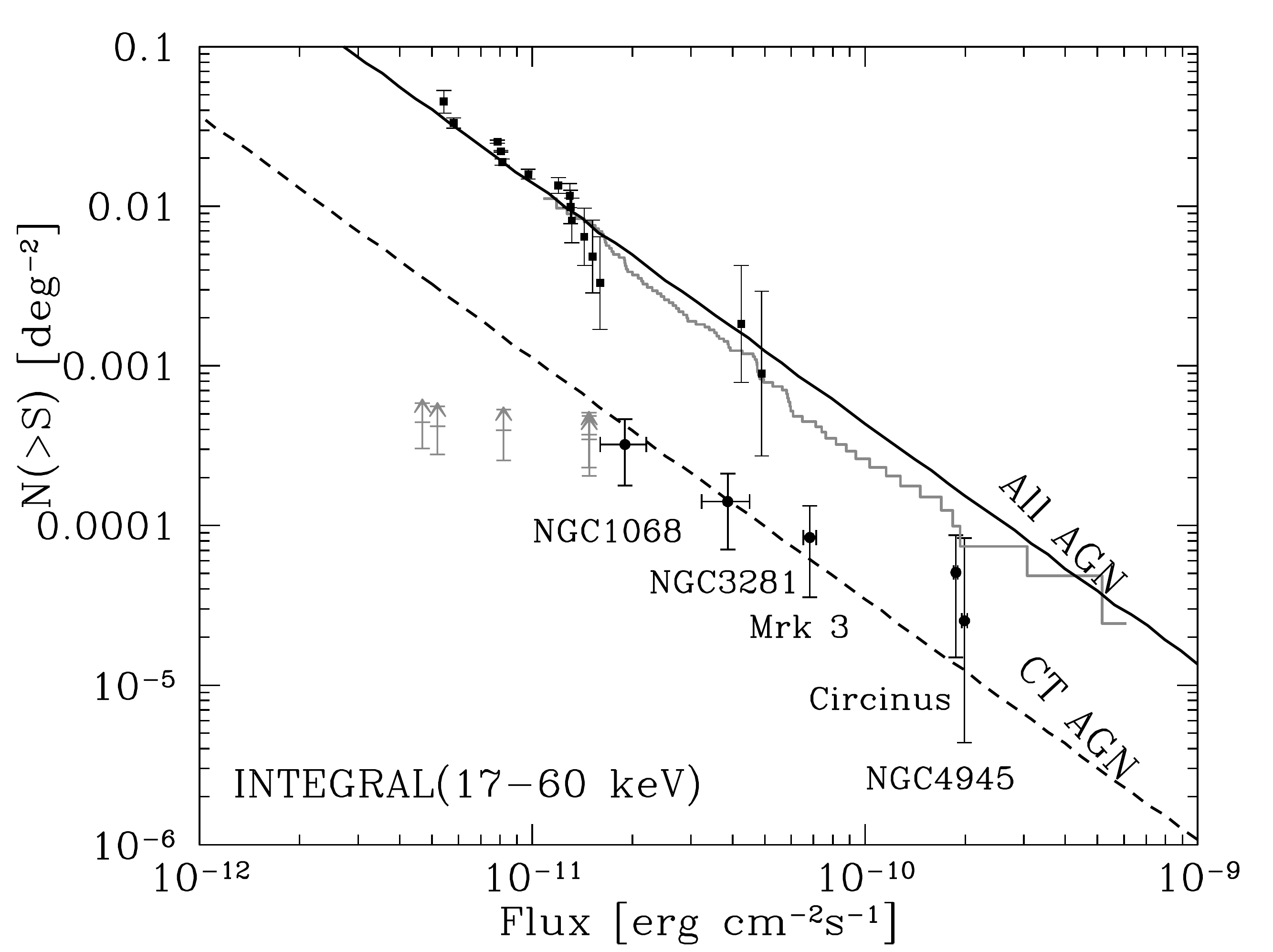}
\vspace{-0.3cm}
\caption{LogN-$\log$S distribution for AGN detected at high energies ($E$$>$10 keV). The {\it gray line} in the {\it left panel} 
shows the AGN in the well-defined Swift/BAT samples in the 14-195 keV band \cite{tueller08}, while the {\it right panel} 
shows the INTEGRAL sources \cite{krivonos07} in the 17-60 keV band. {\it Solid squares} show the 15 sources detected in 
the ultra-deep 3 Msec INTEGRAL observations of the XMM-LSS field. {\it Solid circles} mark the CT AGN detected with Swift 
({\it left panel}) and INTEGRAL ({\it right panel}); the fraction is $\sim$5\%. The {\it black solid lines} show the expected 
AGN $\log$N-$\log$S from the most complete AGN population synthesis model \cite{treister05b}, which at these high fluxes 
has a Euclidean slope. The {\it dashed lines} mark the Euclidean slope normalized to the number of Swift and INTEGRAL 
CT AGN (5\% of the total). The {\it gray} lower limits show the previously-known transmission-dominated AGN with hard 
X-ray observations, not detected in the INTEGRAL or Swift surveys. These are lower  limits since they were selected from 
pointed observations and are thus highly incomplete.}
\label{logn_s_integral}
\end{center}
\end{figure}

Figure~\ref{logn_s_integral} shows the cumulative number counts of AGN, with CT sources shown separately, as a function of 
hard X-ray flux. In order to avoid the necessity of specifying a standard spectrum to convert fluxes to different energy bands, we
show the INTEGRAL and Swift sources separately, but note that there is good agreement (within $\sim$40\%) in the 
normalization between the two distributions if a standard band conversion is assumed. At these high fluxes the slope of 
the $\log$~N-$\log$~S is Euclidean, implying a uniform spatial distribution, as expected given the low redshifts of these sources. 
The number of CT AGN found by these surveys is surprisingly low, compared to the sample of known CT AGN in the local 
Universe, most likely due to the effects of obscuration even at these high energies \cite{treister09b,malizia09,burlon11}. A study of optically-selected local Seyfert 2 
galaxies with hard X-ray information \cite{risaliti99} found 12 CT AGN in a total of 45 Seyfert galaxies. Three were detected by Chandra 
and/or XMM, while the rest are mostly reflection-dominated 
sources, too faint to be detected by either INTEGRAL or Swift even though they are nearby, moderate luminosity AGN. This
suggests that even hard X-ray surveys miss quite a few Compton thick AGN.

\begin{figure}
\begin{center}
\includegraphics[angle=0,scale=0.4]{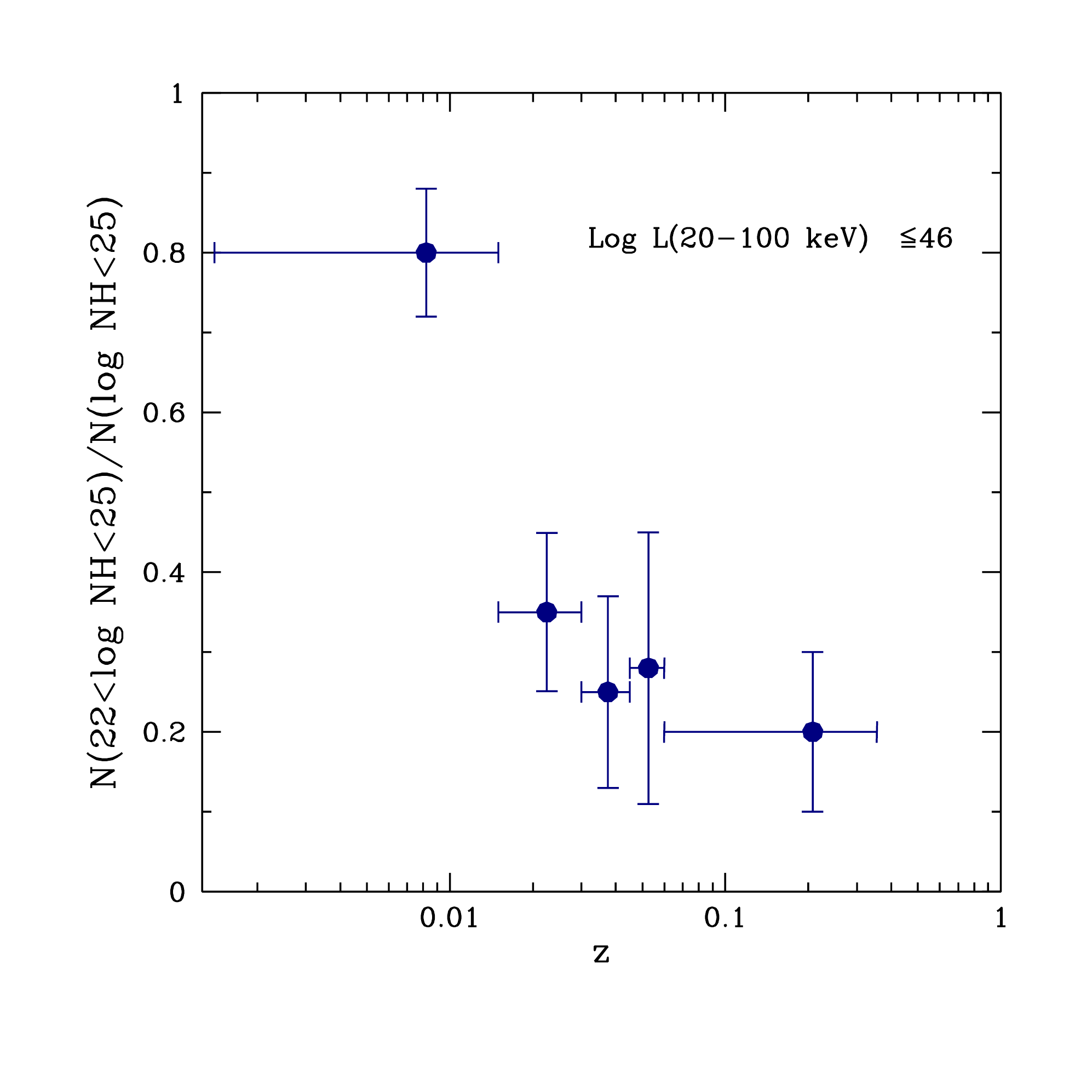}
\vspace{-1.2cm}
\caption{Fraction of obscured AGN ($N_H$$>$10$^{22}$~cm$^{-2}$) as a function of redshift for a sample of AGN selected at
high energies from INTEGRAL observations, as shown by Malizia et al. \cite{malizia09}. All of these sources have X-ray luminosities lower than 
10$^{46}$~erg~s$^{-1}$. The remarkably strong decline in this fraction at $z$$\sim$0.015 is likely due to selection effects and 
not intrinsic to the AGN population. That the fraction of CT AGN in the first bin is $\sim$20\%, in contrast to the $\sim$5\% 
found overall suggests current hard X-ray surveys are not sensitive enough to observe CT AGN beyond $z$$\sim$0.01.
This will improve dramatically with NuSTAR \cite{harrison10}.}
\label{obsc_frac_malizia09}
\end{center}
\end{figure}

The observed fraction of CT AGN in the INTEGRAL and Swift/BAT hard-X-ray selected samples is low, $\sim$5\%. A very 
similar and consistent value, 4.6\%, was recently obtained from a sample of 307 objects detected in the three-years 
all-sky {\it Swift}/BAT survey \cite{burlon11}. This was initially surprising, since previous AGN population synthesis models 
that can explain the XRB used much higher CT fractions of $\sim$15-20\% \cite{treister05b,gilli07}, i.e., factors of 3-4 higher. 
We now know that even observations at these high energies can be affected by obscuration, if the column density is high 
enough. For example, $\sim$50\% of the source flux in the 15-55 keV range can be lost if $\log$N$_H$$>$24.5 
\cite{ghisellini94}. As pointed out by Malizia et al. \cite{malizia09}, and as can be clearly seen in Fig.~\ref{obsc_frac_malizia09}, 
the INTEGRAL all-sky observations, which have a similar flux limit as the {\it Swift}/BAT images, show a steep decline in 
the number of obscured sources, from $\sim$80\% at $z$$<$0.015, to $\sim$20-30\% at higher redshifts. Also, all the CT
AGN in this sample were found at $z$$<$0.015. Hence, these authors concluded that the INTEGRAL 
observations are affected by obscuration at larger distances, and that the intrinsic fraction of CT sources is
$\sim$25\%, as observed at $z$$<$0.015. However, it is worth mentioning that these additional sources,
because of their very high column densities, do not contribute significantly to the XRB --- although they certainly 
contribute to black hole growth.

\begin{figure}
\begin{center}
\includegraphics[angle=270,scale=0.4]{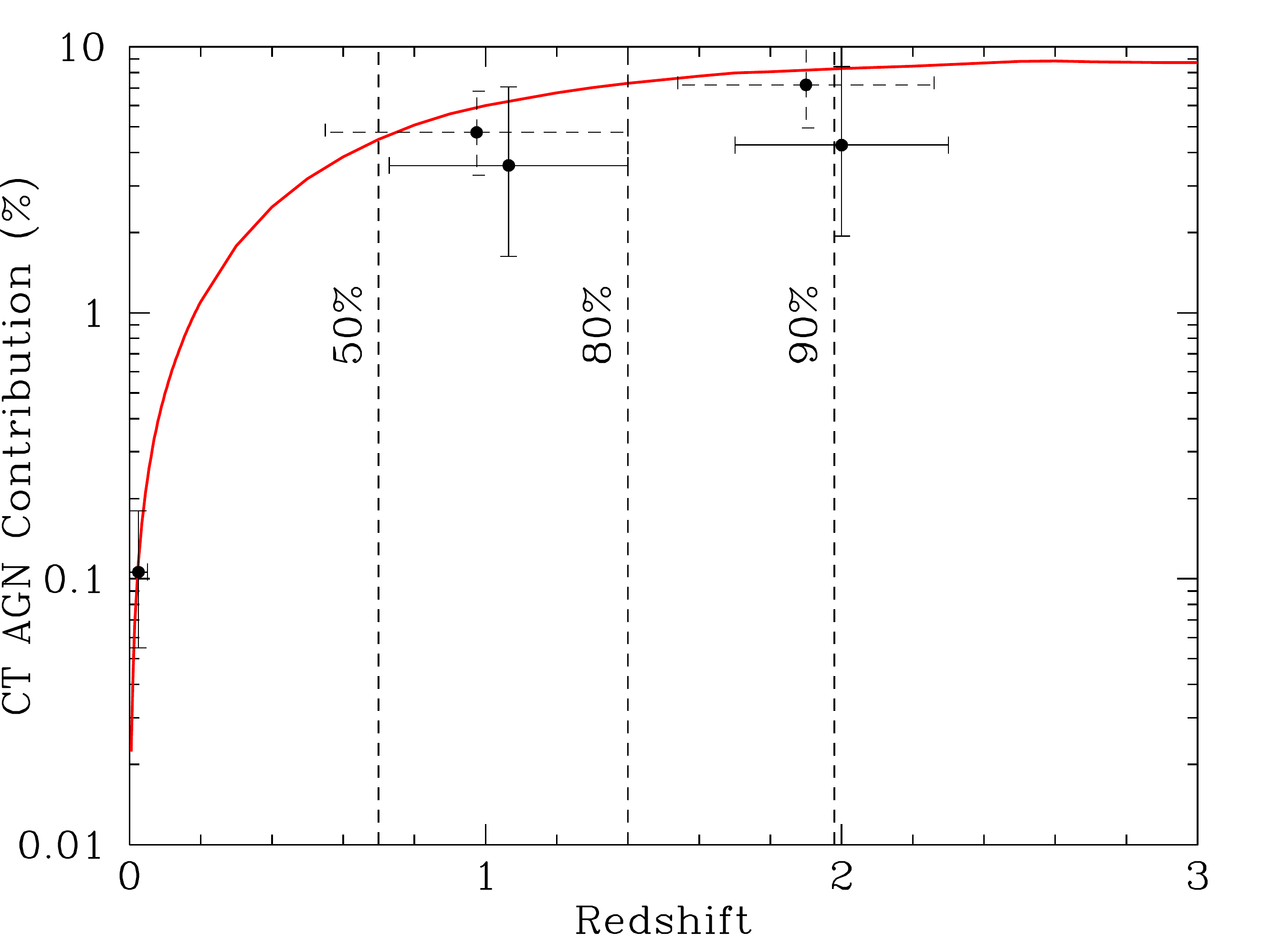}
\vspace{-0.3cm}
\caption{Cumulative fractional contribution of CT AGN to the XRB in the 14-195 keV Swift/BAT band as a function 
of redshift from population synthesis models \cite{treister09b}. As shown by the vertical dashed lines, 50\%, 80\% and 90\% 
of the total CT AGN contribution come from sources at $z$$<$0.7, 1.4 and 2.0, respectively. Only $\sim$1\% of the total 
XRB intensity comes from CT AGN at $z$$>$2. The data point at $z$$\sim$0 corresponds to the contribution
to the XRB by the CT AGN detected by Swift/BAT, while the data points at high redshift were obtained from the 
CT AGN in the CDF-S \cite{tozzi06}. Solid error bars correspond to transmission-dominated sources only, while 
the data points with dashed error bars include all the sources in the sample.}
\label{xrb_frac}
\end{center}
\end{figure}

The cumulative contribution of CT AGN to the XRB, as a function of redshift, determined from population synthesis models, 
is shown in Fig.~\ref{xrb_frac}. As can be seen, the total contribution of CT AGN to the XRB is $\sim$9\%, and about 50\% of 
it comes from sources at $z$$<$0.7. Similarly, only $\sim$2\% of the XRB is provided by CT AGN at $z$$>$1.4,
while CT AGN at $z$$>$2 contribute only $\lsim$1\% to the XRB. Conversely, the 5\% uncertainty in the 
absolute measurement of the XRB intensity translates into an uncertainty of a factor of $\sim$5 in the number 
of CT AGN at $z$$>$2. Hence, the number of CT AGN at high redshift is largely unconstrained by the XRB.

\subsection{Obscured AGN at Intermediate Redshifts (1\lsim z \lsim 3)}

As shown in the previous section, the number of heavily-obscured AGN at intermediate
redshifts, $z$$\gsim$1, is largely unconstrained by the XRB due to model degeneracies, or 
by current X-ray surveys at E$>$10~keV, which do not have the required sensitivity. NuSTAR will change
this situation dramatically. However, for now we are forced to use alternative methods to determine the amount 
of black hole growth occurring in these sources. We explore here two of these techniques, which have been 
particularly successful: X-ray stacking and mid-IR AGN selection.

\begin{figure}
\begin{center}
\vspace{-0.8cm}
\includegraphics[angle=0,scale=0.2]{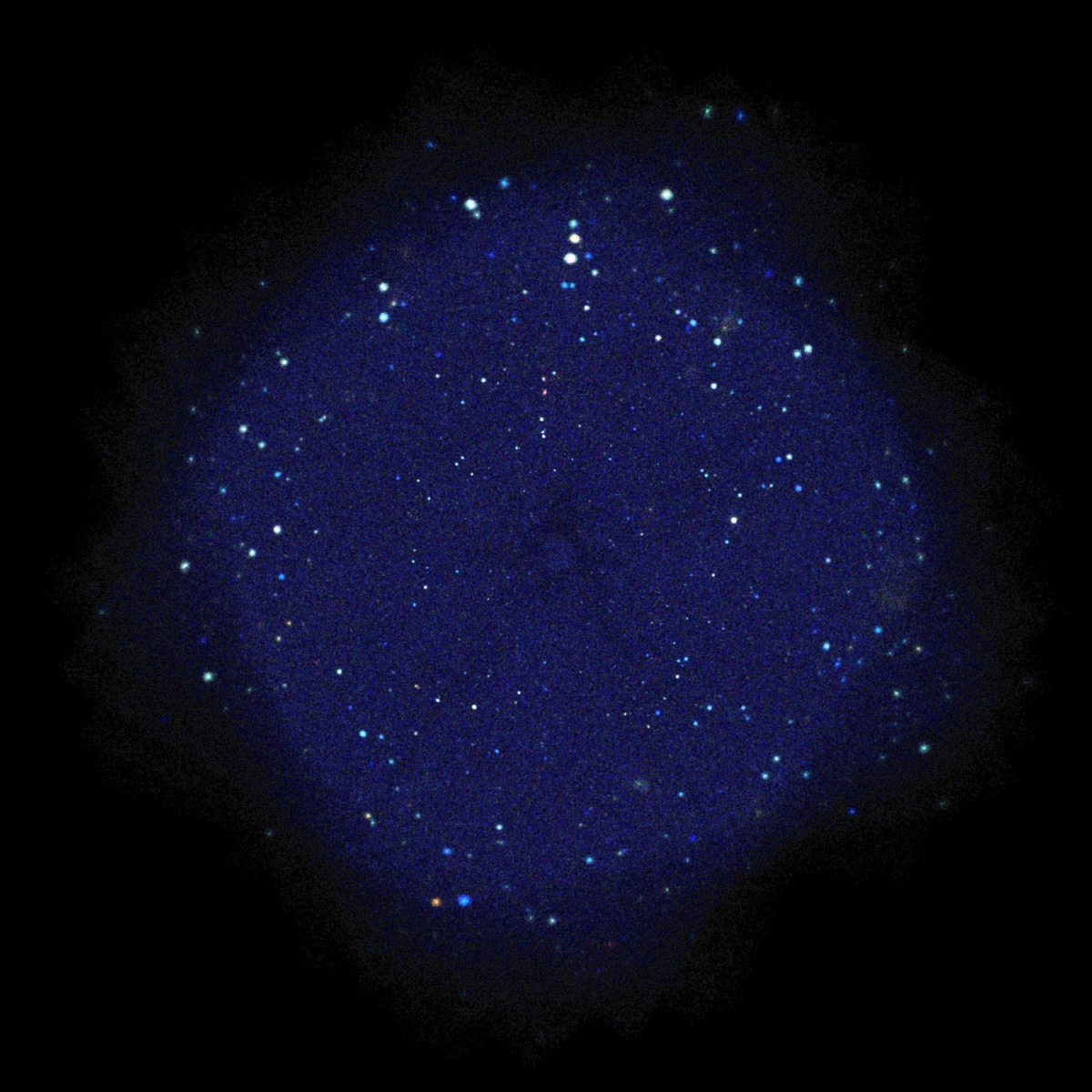}
\caption{4 Msec {\it Chandra} observations of the CDF-S. This is currently the deepest
view of the X-ray sky. There are $\sim$760 individual sources in $\sim$450 arcmin$^2$.
This is a smooth enhanced image corresponding to the full (0.5-8 keV) X-ray band.
Image and data obtained from http://cxc.harvard.edu/cdo/cdfs.html.
}
\label{cdfs}
\end{center}
\end{figure}

Using the deepest available X-ray observations obtained with {\it Chandra}, as in the example shown
in Fig.~\ref{cdfs}, a number of moderately-obscured AGN have been identified at $z$$>$1. The observed-frame
hard X-ray band (2-8 keV band) at these redshifts covers higher rest-frame energies, thus making
them less affected by obscuration. For example, the vast majority of the sources the GOODS fields with high X-ray to 
optical flux ratios are obscured at $z$$\sim$2 \cite{alexander01,barger03a,mainieri05}. Furthermore, from X-ray spectral 
analysis, $\sim$30 CT AGN candidates have been identified in the CDF-S \cite{tozzi06} and CDF-N 
\cite{georgantopoulos09}. However, it is clear that X-ray selection remains highly incomplete for obscured sources 
at these redshifts \cite{treister04}.

\begin{figure}[h!]
\begin{center}
\includegraphics[angle=0,scale=0.4]{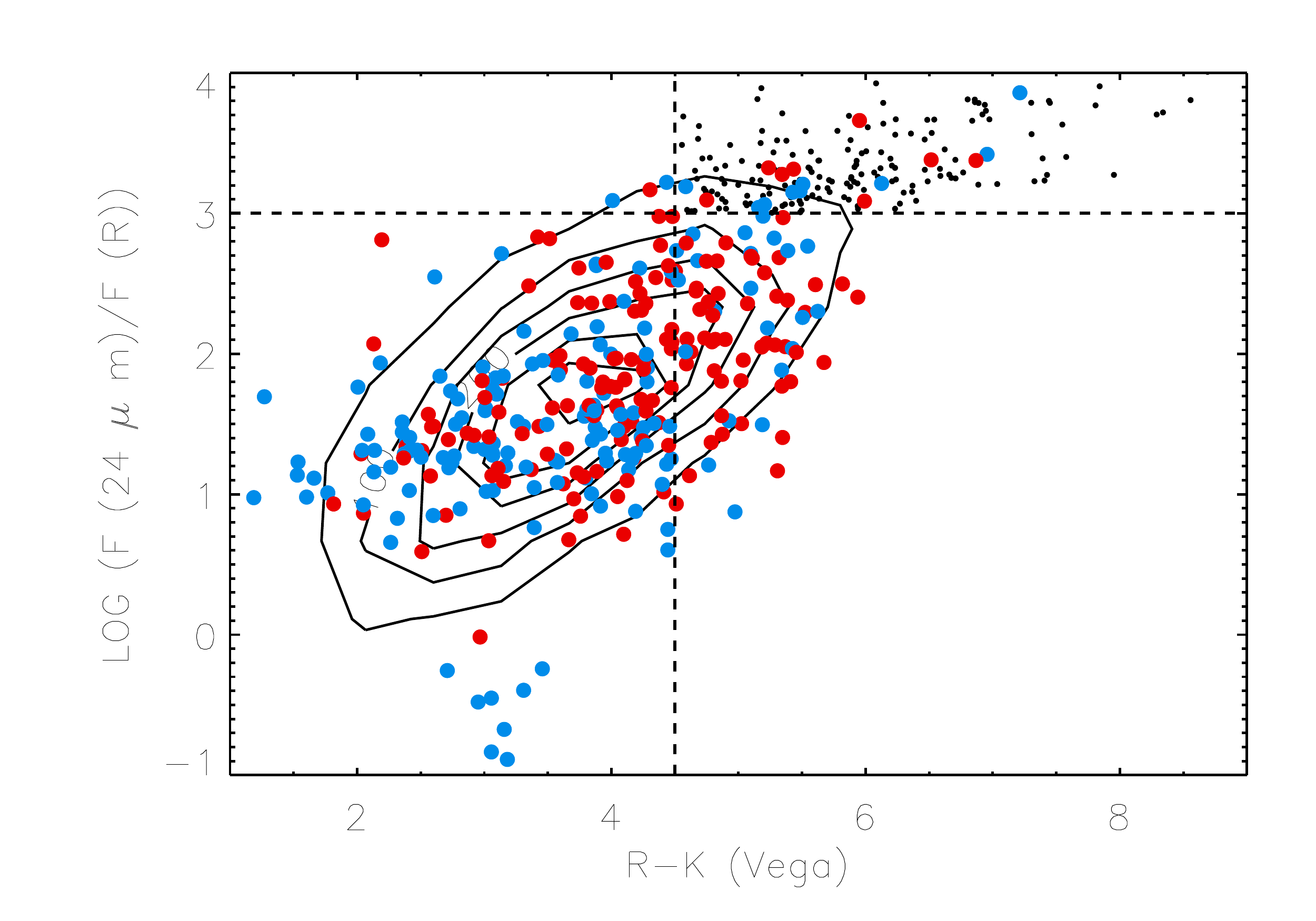}
\vspace{-0.6cm}
\caption{The ratio of 24-micron to $R$-band optical flux as a function of $R$-$K$ color for the sources in the E-CDFS 
field \cite{treister09c}. Contours show the location of all the Spitzer 24~$\mu$m sources. {\it Red points} show the location 
of X-ray sources with hard X-ray spectra (hardness ratio HR$>$-0.3, defined as HR=(H-S)/(H+S) where S and H are the 
background-subtracted counts in the soft and hard bands respectively), while {\it blue points} have HR$<$-0.3. {\it Small 
black points} show the 193 sources in the IR-red excess region that were not detected individually in X-rays. }
\label{dog_sel}
\end{center}
\end{figure}

Because much of the energy absorbed at optical to X-ray wavelengths is later re-emitted in the mid to far-IR, it is expected 
that AGN, in particular the most obscured ones, should be very bright mid-IR sources \cite{treister06a}. Sources having 
mid-IR excesses, relative to their rest-frame optical and UV emission, have been identified as potential CT AGN candidates 
at $z$$\sim$2 \cite{daddi07,fiore08,georgantopoulos08,alexander08,treister09c}. However, because of the strong connection 
between vigorous star formation and AGN activity in the most luminous infrared sources \cite{sanders88}, the 
relative contribution of these two processes remains uncertain and controversial \cite{donley08,pope08,georgakakis10}. 
Significant progress has been made by virtue of deep Spitzer observations, in particular using the 24~$\mu$m band. At 
$z$$\sim$1-2, this emission corresponds to rest-frame wavelengths of $\sim$10~$\mu$m, where the contrast between AGN 
and star formation is largest. In order to look for high-luminosity obscured AGN missed by X-ray observations, Fiore et 
al. \cite{fiore08,fiore09} defined the ``mid-IR excess'' region as $f_{24}$/$f_R$$>$1000 and $R$-$K$$>$4.5 (Vega), as shown 
in Fig.~\ref{dog_sel}. As argued by several authors, the fraction of CT AGN in these infrared-excess samples is very 
high, $>$70\% \cite{fiore08,fiore09,treister09c}. 

\begin{figure}
\begin{center}
\includegraphics[angle=0,scale=0.4]{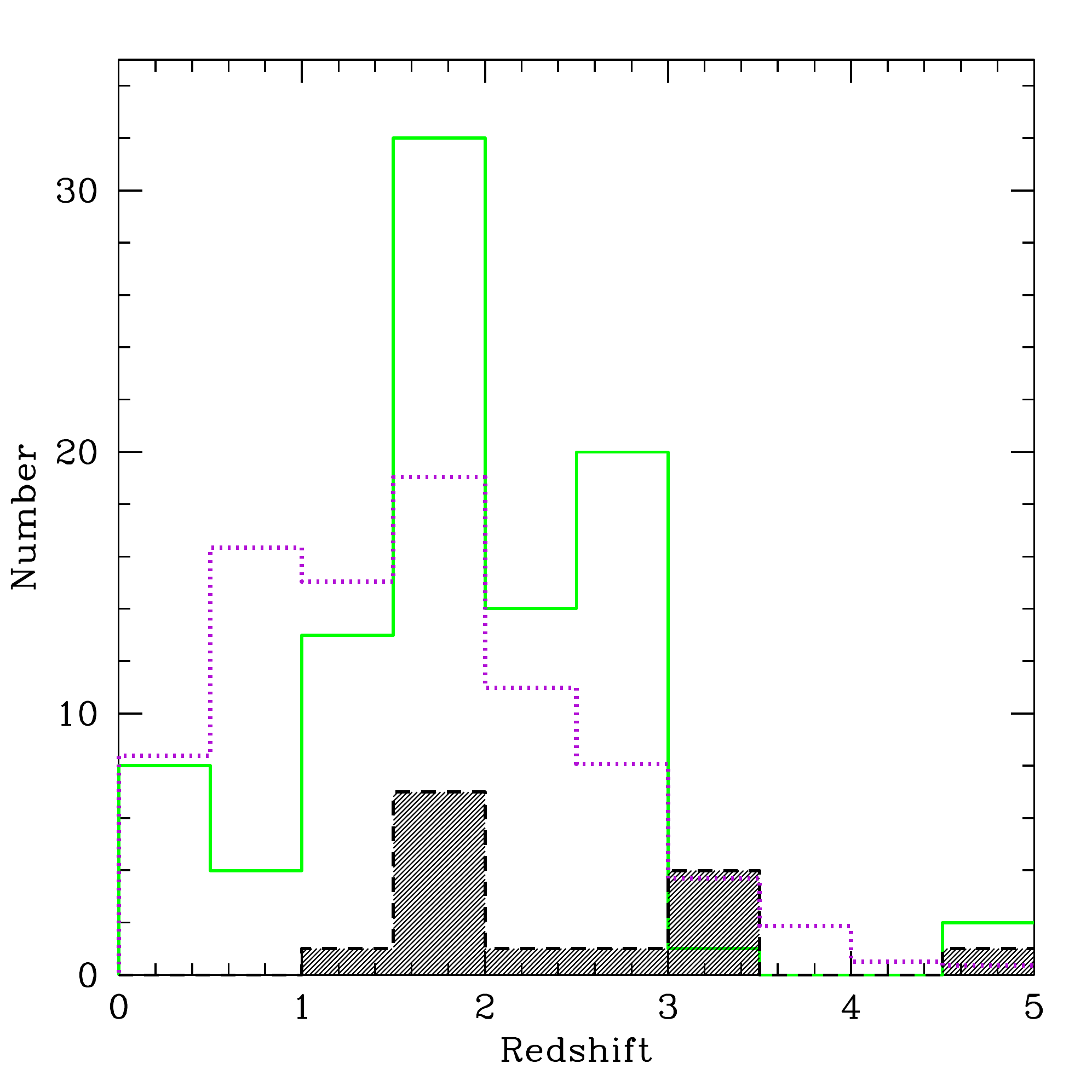}
\vspace{-0.6cm}
\caption{Photometric redshift distribution for 90 sources in the ECDF-S in the mid-IR excess region. The 
{\it solid histogram} shows the distribution for the sources not detected in X-rays, while the {\it dashed hatched histogram} 
considers only the X-ray detected sources. A KS test shows that it is perfectly likely ($\sim$16\%) that these two distributions 
were drawn from the same parent distribution. The {\it dotted histogram} shows the slightly lower redshift distribution (divided 
by 1,000) for all the sources with a 24 $\mu$m detection and a measured photometric redshift in the ECDF-S.}
\label{red_dist_dogs}
\end{center}
\end{figure}

\begin{figure}[h!]
\begin{center}
\includegraphics[angle=270,scale=0.4]{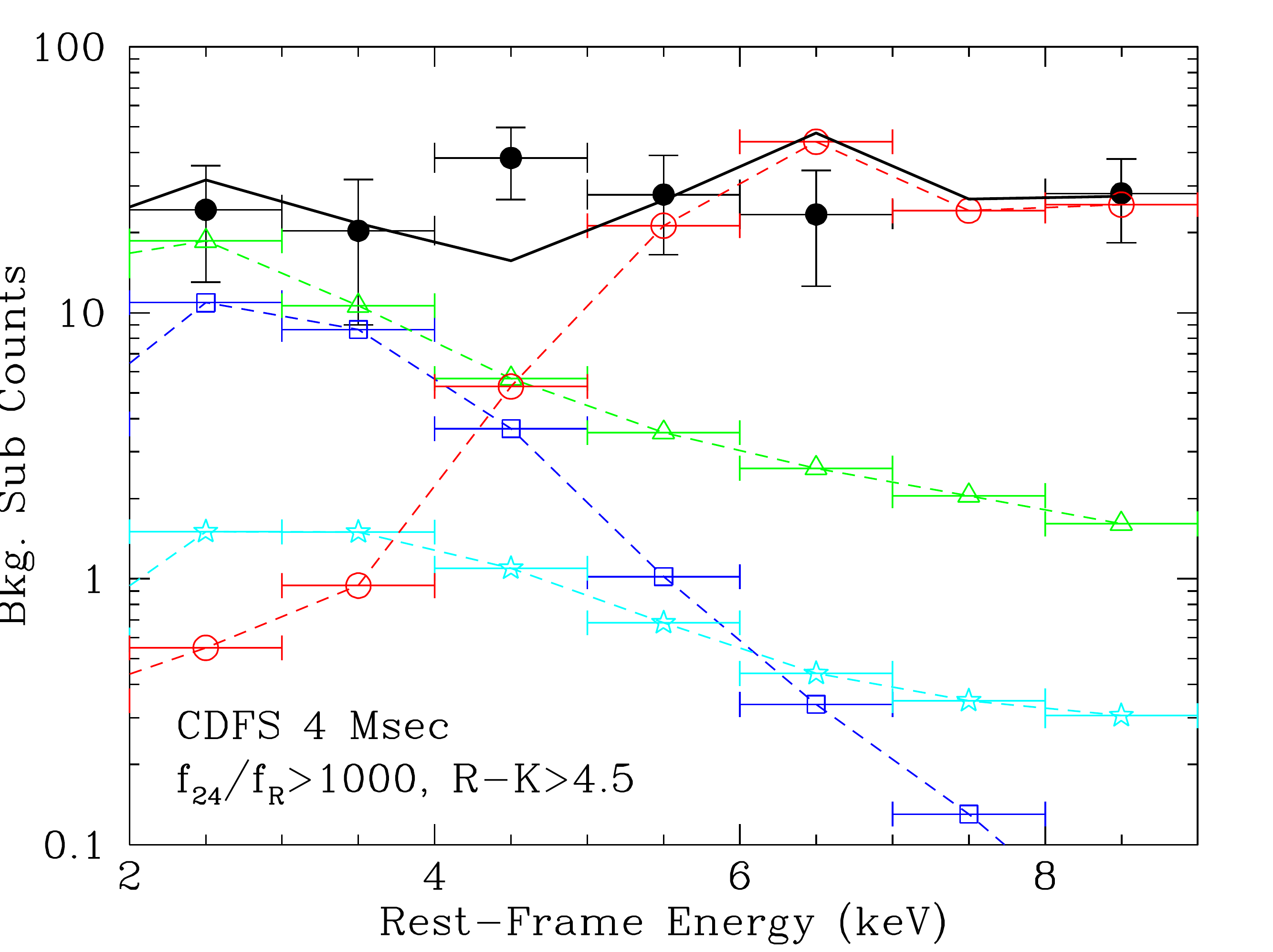}
\caption{Stacked background-subtracted Chandra counts as a function of rest-frame energy, for sources with 
$f_{24}$/$f_R$$>$1000 and $R$-$K$$>$4.5 in the 4 Msec CDF-S field ({\it filled circles}; \cite{treister09c}). The {\it cyan dashed 
lines (stars)} shows the simulated spectra for the high-mass X-ray binary (HMXB) population normalized using the 
relation between star-formation rate and X-ray luminosity \cite{ranalli03}. The {\it blue dashed lines (open squares)} show 
simulated thermal spectra corresponding to a black body with $kT$=0.7 keV. An absorbed AGN spectrum, given by a 
power-law with $\Gamma$=1.9 and a fixed $N_H$=10$^{24}$~cm$^{-2}$, is shown by the {\it red dashed lines (open circles)}. 
In addition, a scattered AGN component, characterized by a 1\% reflection of the underlying unobscured power-law, is 
shown by the {\it green dashed lines (open triangles)}. The resulting summed spectrum ({\it black solid lines}) is in 
very good agreement with the observed counts. The strong detection  in the stacked spectrum at E$>$5 keV, confirms 
the presence of a significant number of heavily-obscured AGN in these IR-excess objects \cite{treister09c}.}
\label{dogs_rest_spec}
\end{center}
\end{figure}

Because sources in the mid-IR excess region are, by definition, very faint at optical wavelengths, it has been very 
difficult to use optical spectroscopy to measure accurate redshifts. Instead, most surveys have had to rely on (hopefully) 
accurate photometric redshifts \cite{salvato09,cardamone10}. The distribution of photometric redshifts for the sources in 
the mid-IR excess region in the ECDF-S is shown in Fig.~\ref{red_dist_dogs}; most mid-IR excess sources, have 
1$<$$z$$<$3. While the majority of these sources are not detected in X-rays, a significant signal is found in X-ray 
stacks \cite{fiore08,fiore09,treister09c}. As shown in Fig.~\ref{dogs_rest_spec}, the strong stacked detection 
at $E$$>$5~keV clearly indicates the presence of a large number of heavily-obscured AGN in this infrared-excess sub-sample.
Specifically, Treister et al. \cite{treister09c} reported that heavily-obscured AGN were $\sim$80-90\% of the mid-IR-excess
sources in the ECDF-S. A similarly high fraction of $\sim$80\% was found in the CDF-S \cite{fiore08} and other 
fields \cite{fiore09}. Optical spectral fitting of these sources indicates evidence for substantial young stellar populations, 
younger than 100 Myrs \cite{treister09c}. This suggests that these sources are simultaneously experiencing significant 
star-formation and heavily-obscured AGN activity. The best-fit stellar masses for ECDF-S infrared-excess sources range 
between 10$^{9}$ and 10$^{12}$$M_{\odot}$ with a median stellar mass of $\sim$10$^{11}$$M_\odot$ \cite{treister09c}. 
Hence, in general these are very massive galaxies.

\begin{figure}[h!]
\begin{center}
\includegraphics[angle=0,scale=0.4]{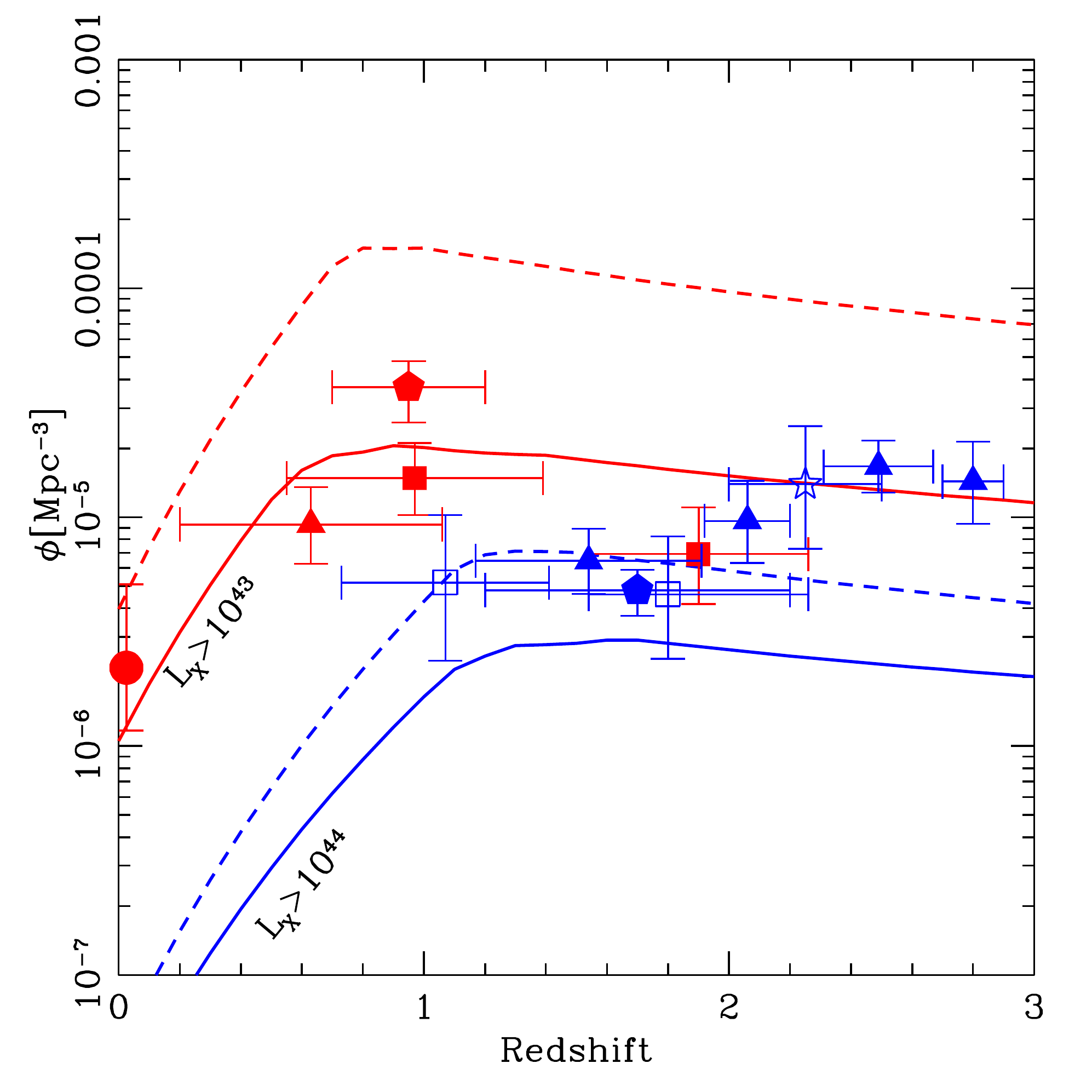}
\vspace{-0.6cm}
\caption{Space density of Compton thick AGN as a function of redshift, as published by Treister et al. \cite{treister09c}. 
{\it Filled triangles} show the ECDF-S results \cite{treister09c}. {\it Squares}: X-ray selected sources in the CDF-S 
\cite{tozzi06}. {\it Star}: Measurement obtained using mid-IR spectroscopy \cite{alexander08}. {\it Pentagons}: Values 
obtained using mid-IR excess sources in COSMOS \cite{fiore09}. {\it Solid lines} show the expected space density of 
Compton thick AGN from the luminosity function of Yencho et al. \cite{yencho09}, with the overall normalization fixed to the 
results of the INTEGRAL and Swift/BAT surveys \cite{treister09b}, while the {\it dashed lines} show the expectations based 
on the luminosity function of Della Ceca et al. \cite{dellaceca08a}. {\it Red symbols} show measurements and expectations 
for $L_X$$>$10$^{43}$~erg~s$^{-1}$ sources, while the {\it blue symbols} are for $L_X$$>$10$^{44}$~erg~s$^{-1}$. While 
for the lower luminosity sources a good agreement is found between observations and expectations, higher luminosity 
sources at $z$$>$1.8 lie well above the luminosity function.}
\label{com_dens}
\end{center}
\end{figure}

In order to study in more detail the evolution of the CT AGN space density, in Fig.~\ref{com_dens} we present the existing 
measurements of the CT AGN space density as a function of redshift. Reasonable agreement, in particular at $z$$<$1, is 
found between both observed values and existing hard X-ray luminosity functions and evolution \cite{yencho09}. However, at 
$z$$>$1.7 and high luminosity, a clear discrepancy is found. Treister et al. \cite{treister09b}, concluded that this difference of 
a factor of 2--3 could be due either to incompleteness in the Swift/BAT and INTEGRAL CT AGN samples at $z$=0 used to fix 
the luminosity function normalization (because reflection-dominated AGN are missed) or to contamination by other types of 
sources in the observed values at high redshifts. However, after adding the measurements obtained using the infrared-excess 
sources in the ECDF-S, it appears not only that the systematic difference is still present but perhaps more importantly that there 
is a strong  increase in the number of CT AGN from $z$$\simeq$1.7 to 2.4. This is not described by any existing luminosity 
function. It is unlikely that this evolution is due to selection effects, as results from different fully independent surveys and 
selection techniques are combined in Fig.~\ref{com_dens}, namely X-ray selected sources \cite{tozzi06}, 24~$\mu$m-selected 
sources \cite{fiore09,treister09c}, and a sample of CT AGN found using mid-IR spectroscopy \cite{alexander08}. This result can 
be interpreted in the context of galaxy evolution models \cite{hopkins08}, where quasar activity is driven by galaxy mergers and 
the supermassive black hole is initially completely surrounded by dust, before radiation pressure removes it and a ``classical'' 
unobscured quasar is visible \cite{treister10}.

\begin{figure}[h!]
\begin{center}
\includegraphics[angle=0,scale=0.4]{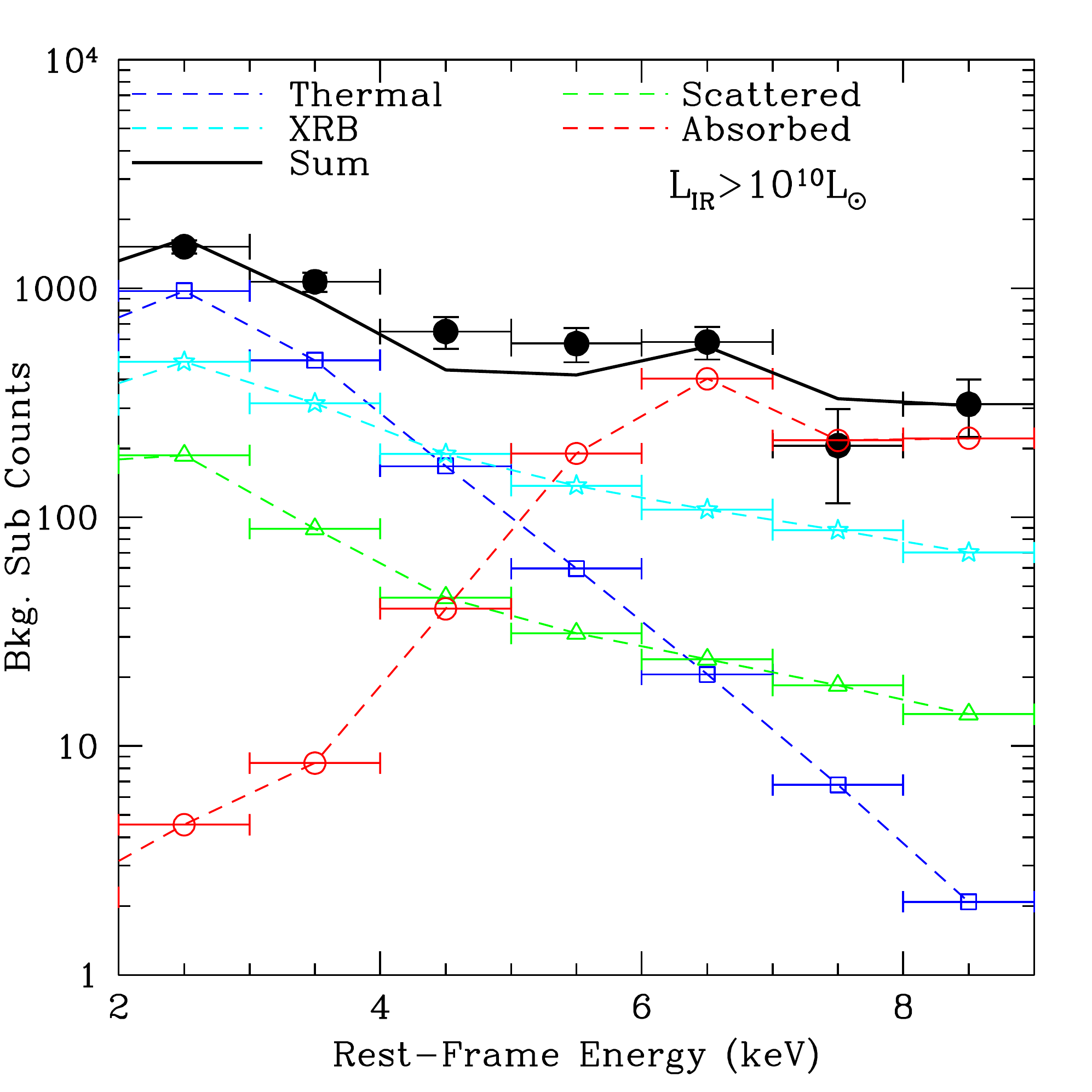}
\includegraphics[angle=0,scale=0.4]{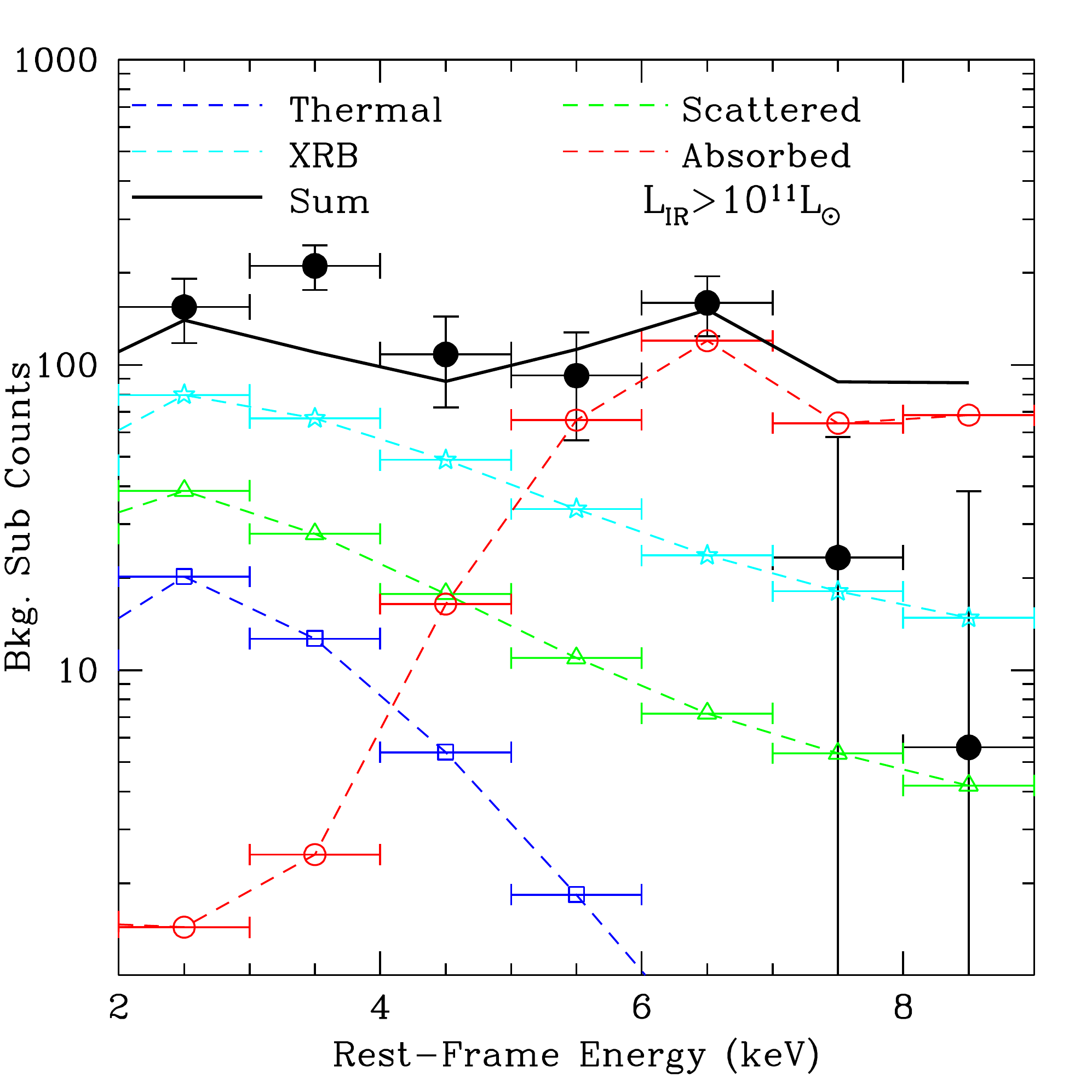}
\caption{Stacked background-subtracted Chandra counts as a function of rest-frame energy for sources with low
({\it left panel}) and high ({\it right panel}) IR luminosity ({\it black filled circles}). The {\it cyan dashed lines (stars)}, 
{\it blue dashed lines (open squares)}, {\it red dashed lines (open circles)} and {\it green dashed lines (open triangles)} show 
the same model components as in Fig.~\ref{dogs_rest_spec}. The resulting summed spectra ({\it black solid lines}) 
are in very good agreement with the observed counts. The strong detection in the stacked spectrum at E$>$5 keV, in particular 
at the higher IR luminosities, confirms the presence of a significant number of heavily-obscured AGN in these samples.}
\label{ir_stack_spec}
\end{center}
\end{figure}

Several groups \cite{kartaltepe10} have found that the fraction of galaxies containing an AGN is a strong function of their IR 
luminosity. In Fig.~\ref{ir_stack_spec} we present the stacked spectra for the sources in the CDF-S, grouped in bins of IR 
luminosity \cite{treister10b}. We can see by comparing these spectra that the relative emission at E$\gtrsim$5 keV, where 
we expect the AGN emission to dominate even for heavily-obscured sources, changes with IR luminosity. In other words, there 
is a clear trend, with stronger high energy X-ray emission at increasing IR luminosity. The spectra shown in 
Fig.~\ref{ir_stack_spec} cannot be directly interpreted, as the detector-plus-telescope response information is lost after 
the conversion to rest-frame energy and stacking. Hence, simulations assuming different intrinsic X-ray spectra have to be 
used in order to constrain the nature of the sources dominating the co-added signal.

The observed stacked spectral shape cannot be explained by any plausible starburst spectrum. An AGN 
component dominating at E$>$5~keV, is required. The average intrinsic rest-frame 2-10 keV AGN luminosity 
needed to explain the observed spectrum, assuming that every source in the sample contains an AGN of the same 
luminosity, is 6$\times$10$^{42}$~erg~s$^{-1}$ for sources with $L_{IR}$$>$10$^{11}$$L_\odot$, 
3$\times$10$^{42}$~erg~s$^{-1}$ for sources with $L_{IR}$$>$5$\times$10$^{10}$$L_\odot$, 
5$\times$10$^{41}$~erg~s$^{-1}$ for 5$\times$10$^{10}$$L_\odot$$>$$L_{IR}$$>$10$^{10}$$L_\odot$ and 
7$\times$10$^{41}$~erg~s$^{-1}$ for $L_{IR}$$>$10$^{10}$$L_\odot$. All of these are (intrinsically) very low-luminosity 
AGN; even if there is a range, it is extremely unlikely to include high-luminosity quasars like those discussed in previous 
stacking papers. This is not too surprising, actually, because the surveyed volume (even to high redshift) is small, so rare
objects like high-luminosity quasars do not appear. If the heavily-obscured AGN in these stacked samples have the same 
median intrinsic luminosity as the X-ray detected sources with similar IR luminosities, this would indicate that 15\% of the 
galaxies with $L_{IR}$$>$10$^{11}$$L_\odot$ contain a heavily-obscured AGN. This fraction is $\sim$10\% in the 
$L_{IR}$$>$5$\times$10$^{10}$$L_\odot$ and 5$\times$10$^{10}$$L_\odot$$>$$L_{IR}$$>$10$^{10}$$L_\odot$ samples. 
For sources with $L_{IR}$$>$10$^{10}$$L_\odot$ this fraction is $<$5\%. This extra AGN activity (in addition to the
X-ray detected sources) can account for $\sim$22\% of the total black hole accretion. Adding this to the obscured black 
hole growth in X-ray detected AGN \cite{luo08}, we confirm that most of this growth, $\sim$70\%, is significantly obscured 
and missed by even the deepest X-ray surveys \cite{treister04,treister10}.

\subsection{Obscured AGN at High redshifts, $z$$>$3}

\begin{figure}
\begin{center}
\includegraphics[angle=0,scale=0.4]{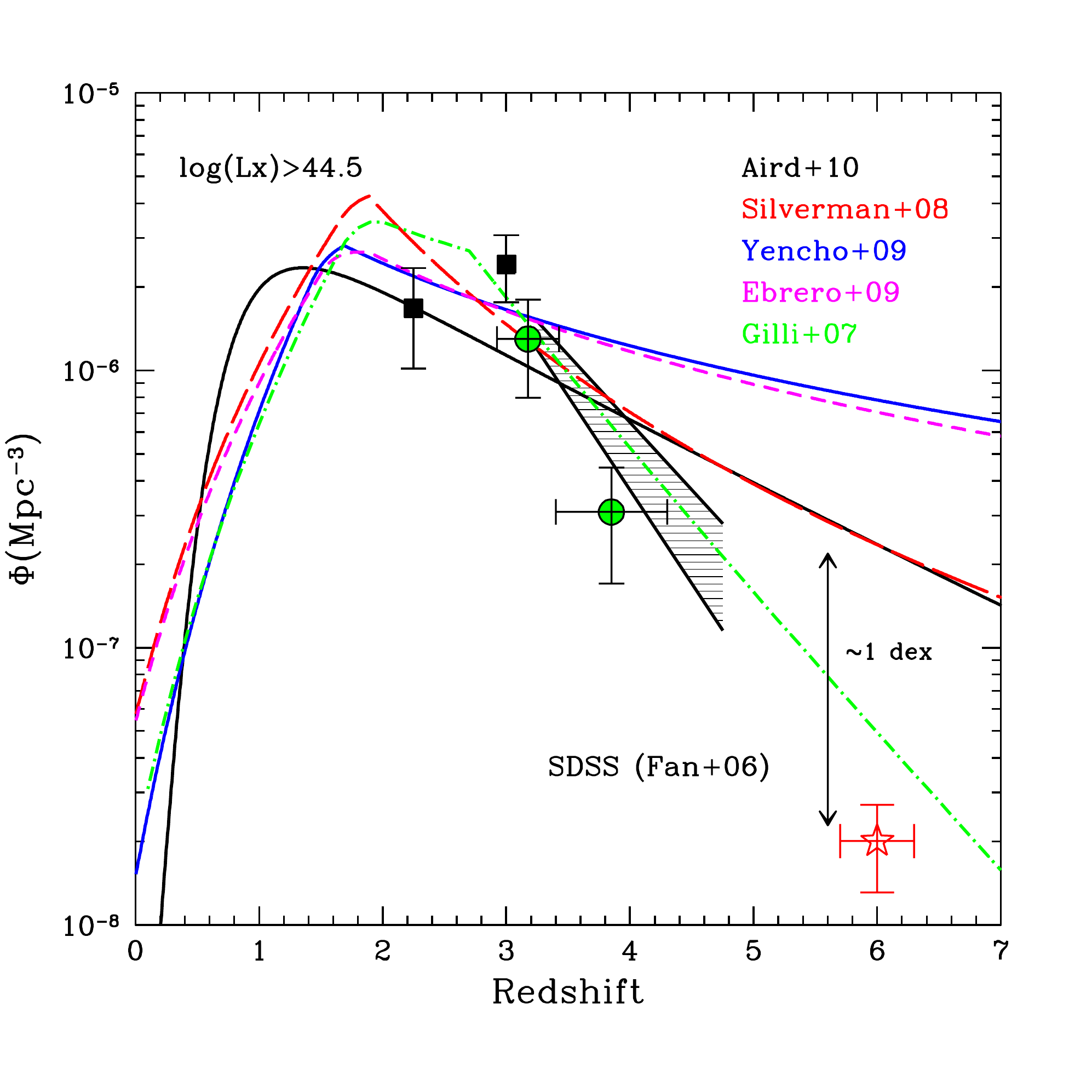}
\caption{Number density of X-ray selected AGN as a function of redshift for sources with $\log$$L_{2-10}$$>$44.5 erg~s$^{-1}$,
as published by Brusa et al. \cite{brusa11}. {\it Lines} show AGN luminosity functions 
\cite{gilli07,silverman08b,ebrero09,yencho09,aird10} while individual measurements are shown by green circles 
\cite{brusa09} and black squares \cite{aird10}. The {\it red star} at $z$=6 was obtained from the optical luminosity 
function assuming no evolution in $\alpha_{ox}$. The {\it black shaded area} shows the observations of optically-bright 
SDSS QSOs \cite{fan06}.}
\label{dens_red_brusa11}
\end{center}
\end{figure}

As mentioned in \S\ref{unobscured}, most measurements of black hole accretion at high redshift, $z$$>$3, come 
from optical observations of unobscured sources. This is not only because obscured sources are obviously fainter at most
wavelengths, but also because large areas have to be covered in order to survey a significant volume at high redshifts. As 
can be seen in Fig.~\ref{dens_red_brusa11}, there is a clear decline in the number of luminous quasars at $z$$>$2,
although the decline is shallower for X-rays compared to optical surveys, since X-ray selection is less biased 
\cite{ebrero09,yencho09}. However, it is important to point out that: ($i$) these results are limited to the highest 
luminosity sources, $\log$$L_X$$>$44.5~erg~s$^{-1}$, which do not represent the average AGN and do not contribute 
much to the extragalactic XRB \cite{treister09b}, and ($ii$) only relatively unobscured sources are included. In particular, 
heavily obscured, Compton-thick, AGN are systematically underrepresented in these surveys. As we will describe, these 
missing populations can have a significant impact in our understanding of cosmic supermassive black hole growth.

In order to search for the presence of growing supermassive black holes in young galaxies, Treister et al. \cite{treister11} 
stacked X-ray images of $z$$>$6 galaxy candidates selected based on the optical and near-IR dropout techniques, selected
from a sample of 197 galaxies, 151 in the CDF-S and 46 in the CDF-N from Bouwens et al. \cite{bouwens06}. Using the
4 Msec Chandra observations of the CDF-S and the 2 Msec data available on the CDF-N, this corresponds to a total exposure 
time of $\sim$7$\times$10$^8$ seconds ($\sim$23 years). Significant detections in both the soft and hard X-ray bands, were 
obtained. However, these detections have recently been questioned (after the submission of this review) by several authors \cite{fiore11,willott11,cowie11} due
to a possible bias in the background subtraction technique used by Treister et al. While a full discussion is clearly beyond the scope of this
paper, we note that a full analysis using an optimal weighting scheme (as in \cite{treister11}), as well as considering the effects of faint,
undetected, sources in the background, remains to be done. At a minimum, the results presented below can be considered as upper
limits.

\begin{figure}
\begin{center}
\includegraphics[angle=0,scale=0.4]{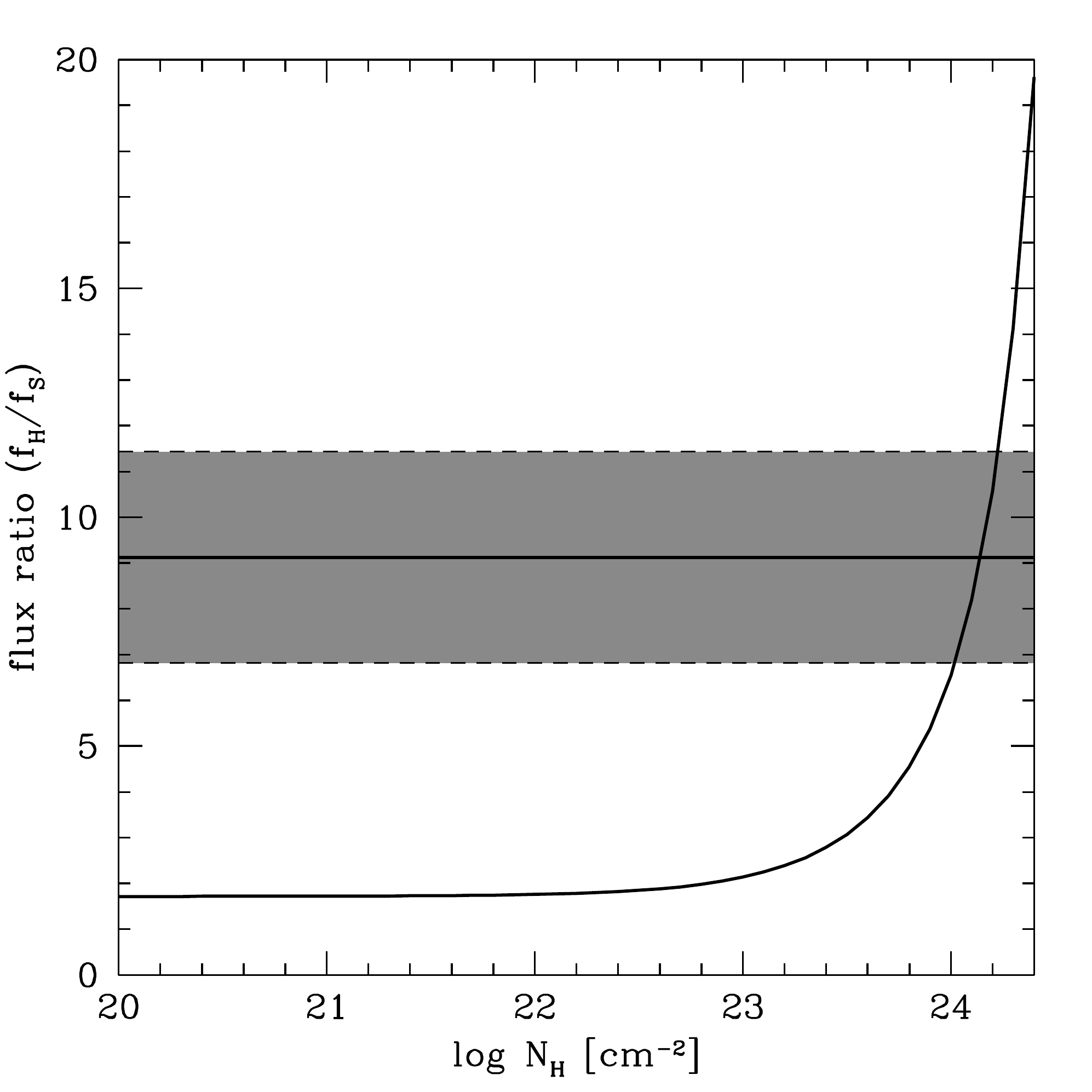}
\caption{Expected ratio of the observed-frame hard to soft flux as a function of obscuring neutral Hydrogen column 
density ($N_H$). The {\it black solid line} was derived assuming an intrinsic power-law spectrum with slope $\Gamma$=1.9 
and photoelectric absorption. The gray zone shows the measured ratio for the stack of galaxies at $z$$\simeq$6 and the 
$\pm$1 standard deviation limits. A column density of N$_H$$\simeq$10$^{24}$~cm$^{-2}$, i.e. Compton-thick obscuration, 
is required to explain the observed hard to soft X-ray flux ratio \cite{treister11}.}
\label{fratio_nh}
\end{center}
\end{figure}

The corresponding average rest-frame 2-10 keV luminosity, derived from the observed-frame hard band, is 
6.8$\times$10$^{42}$~erg~s$^{-1}$. Since none of these sources were individually detected in X-rays,  at least 30\% of 
the galaxies in this sample likely contain an AGN \cite{treister11}. Furthermore, there is a factor $\sim$9 difference between the 
fluxes measured in the observed-frame soft and hard bands. The only explanation for this relatively large flux ratio in the hard 
to soft bands is very high levels of obscuration. As can be seen in Fig.~\ref{fratio_nh}, at $z$$\sim$6, a minimum column 
density of N$_H$$\simeq$10$^{24}$~cm$^{-2}$, i.e. Compton-thick obscuration, is required. Given that this ratio is observed 
in a stacked X-ray spectrum, this implies that there are very few sources with significantly lower levels of obscuration, which 
in turn means that these sources must be nearly Compton-thick along most directions ($\sim$4$\pi$ obscuration). Similar 
sources have also been observed in the local Universe \cite{ueda07} but appear to be rare. Furthermore, for $z$$\lsim$3 
we know that the fraction of obscured AGN increases with decreasing luminosity \cite{ueda03,lafranca05,sazonov07} and 
increasing redshift \cite{treister06b,ballantyne06}. Hence, it is not entirely surprising that the 
sources studied here, given their low luminosities and high redshifts, are heavily obscured. In fact, the discovery of a 
Compton-thick AGN at $z$$\sim$5 selected using the drop-out technique has been recently reported \cite{gilli11}.

This relatively large number of growing supermassive black holes at $z$$\gsim$6 is contrary to the picture obtained from 
optical observations of high-luminosity quasars at similar distances, which are much rarer \cite{fan06,willott10b}. This is 
particularly important in our understanding of the early hydrogen re-ionization, which can be either due to young stars and/or 
growing supermassive black holes \cite{volonteri09}. The results presented by Treister et al. \cite{treister11} show that while 
growing supermassive black holes do not contribute much to hydrogen re-ionization, this is not because their numbers drop 
steeply at $z$$>$4 as previously suggested \cite{Faucher2008}, but because large amounts of obscuration found in these 
sources imply that UV and soft X-rays do not escape.

\section{The Cosmic History of Black Hole Accretion}

Direct black hole mass measurements, either through stellar or gas dynamics, are available for only a few nearby galaxies. 
However, thanks to the tight correlation between mass of the supermassive black hole and other properties such as velocity 
dispersion and others, it has been possible to estimate the black hole mass function at $z$$\simeq$0 
\cite{salucci99,yu02,marconi04,shankar04}. This is commonly done starting from the observed galaxy luminosity or velocity 
function and assuming either a constant black hole to stellar mass ratio \cite{salucci99} or the M-$\sigma$ relation 
\cite{marconi04}.  Both the overall shape of the black hole mass function and the integrated black hole mass density, which 
can only be computed at $z$$\simeq$0, can be used to infer properties of the AGN population. This was first used in the 
so-called ``Soltan's argument'' \cite{soltan82}, which says that the intrinsic bolometric AGN luminosity, $L$, is directly linked to 
the amount of mass accreted by the black hole, $ \dot{M}_{acc}$:

\[
L=\varepsilon \dot{M}_{acc}c^2,
\]

\noindent where $\varepsilon$ is the accretion efficiency and $c$ is the speed of light. A typical value
assumed for the efficiency is $\sim$10\% \cite{soltan82,marconi04}.

Recent comparisons of the black hole mass function to the distribution inferred from the observed AGN luminosity indicate that 
the average efficiency is 8\%, the Eddington ratio is $\sim$50\%, and the average lifetime of the visible AGN phase 
is $\sim$10$^8$ years \cite{marconi04,shankar04}. By studying the black hole mass distribution at the high mass end, 
M$>$10$^9$M$_\odot$, Natarajan \& Treister \cite{natarajan09} found that the observed number of ultra-massive black holes 
is significantly lower than the number density inferred from the AGN hard X-ray luminosity function. They concluded that this 
is evidence for an upper limit to the black hole mass, which can be explained by the presence of a self-regulation 
mechanism.

The observed black hole mass density at $z$$\simeq$0 obtained by integrating the black hole mass function, 
ranges from 2.9$\times$10$^5$ \cite{yu02} to 4.6$^{+1.9}_{-1.4}$$\times$10$^5$~M$_\odot$Mpc$^{-3}$ \cite{marconi04}; 
most recently, Shankar et al. \cite{shankar09} found 3.2-5.4 $\times$10$^5$~M$_\odot$Mpc$^{-3}$. For comparison, 
integrating the AGN hard X-ray LF, including the number of Compton-thick AGN constrained by INTEGRAL and Swift/BAT 
observations, Treister et al. \cite{treister09b} obtained a value of 4.5$\times$ 10$^5$ M$_\odot$~Mpc$^{-3}$, perfectly 
consistent with the observed value, indicating that at least locally, X-ray detected AGN can account for most or all of the 
black hole growth.

The black hole mass function can be measured observationally for unobscured, high luminosity AGN at higher redshift, 
taking advantage of the known correlation between black hole mass and observational quantities such as luminosity and 
emission line width \cite{vestergaard06}. These correlations are calibrated using more direct black hole measurements, 
available for a few, mostly local, sources \cite{blandford82,peterson93}. The large number of unobscured quasars with 
optical spectroscopy provided by the SDSS and other optical surveys have been very useful in determining the black hole 
mass function up to high redshifts, \cite{mclure04,vestergaard04,kelly10}, and for lower luminosity sources using deeper 
surveys such as the AGN and Galaxy Evolution Survey (AGES; \cite{kollmeier06}) and COSMOS \cite{trump09}. Using 
these black hole mass functions as a function of redshift as constraints, recently Natarajan \& Volonteri \cite{natarajan11} 
concluded that the observational data are inconsistent with the hypothesis that these black holes are created as the 
remnants of population III stars. Instead, they argue massive seeding models are required \cite{lodato07}. 

While a clear picture of the history of black hole growth is emerging, significant uncertainties still remain. In particular, 
while the spectral shape and intensity of the extragalactic X-ray background have been used to constrain the AGN 
population, the number of heavily obscured accreting supermassive black holes beyond $z$$\sim$1 is not properly bounded. 
Infrared and deep X-ray selection methods have been useful in that sense, but have not provided a final answer, due to
confusion with star-forming galaxies in the infrared and the effects of obscuration in X-rays. At higher redshifts, the situation 
is even more unclear, and only a few, very rare, high luminosity quasars are known. Unless high-redshift AGN luminosity
functions are pathological, these extreme sources do not represent the typical growing black holes in the early Universe. As 
a consequence, and in spite of recent advances \cite{treister11,natarajan11}, the formation mechanism for the first black 
holes in the Universe is still unknown.

\section{Prospects}

Scheduled for launch in February 2012, NuSTAR will be the first focusing high energy ($E$=5-80 keV) X-ray mission, reaching 
flux limits $\sim$100 times fainter than {\it INTEGRAL} or {\it Swift}/BAT observations and comparable to {\it Chandra} 
and {\it XMM-Newton} at lower energies. During the first two years of operations, NuSTAR will likely observe, as part of 
the guaranteed time program, two extragalactic fields: the ECDF-S and the central 1 deg$^2$ part of COSMOS, for a total of 
3.1 Msec each. These deep high-energy observations will enable to obtain a nearly complete AGN survey, including 
heavily-obscured Compton-thick sources, up to $z$$\sim$1.5 \citep{ballantyne11}. A similar mission, ASTRO-H 
\cite{takahashi10}, will be launched by Japan in 2014. Both missions will provide angular resolutions \lsim1$'$, which 
in combination with observations at longer wavelengths will allow for the detection and identification of most growing 
supermassive black holes at $z$$\sim$1. 

There is little doubt that the Atacama Large Millimeter Array (ALMA) will revolutionize our understanding of galaxy evolution. 
Sources of mm and sub-mm emission traced by ALMA includes thermal emission of the warm/cold dust, which traces star 
formation, synchrotron radiation associated with relativistic particles and free-free radiation from HII regions. In particular, CO 
rotational transition lines have been used to trace the spatial distribution, kinematics, temperature and mass 
of the molecular gas \cite{yao03}. The  sensitivity of ALMA will allow for the detection of luminous IR galaxies 
($L_{\rm IR}$$>$10$^{11}$$L_\odot$), which have been found to often host a heavily-obscured AGN \cite{treister10b}, up 
to $z$$\sim$10. Furthermore, with ALMA it will be possible to study separately the molecular dust surrounding the central 
black hole and those in star forming regions in the host galaxy. Due to their limited sensitivity and relatively bad angular 
resolution, currently-available mm/sub-mm telescopes are not ideal to study star-forming regions even in nearby galaxies. 
This will dramatically change thanks to ALMA, which will have orders of magnitude better sensitivity and HST-like angular 
resolution. The first call for ALMA observations was released on March 31, 2011 for observations starting on September 30, 
2011. It is expected that the complete array will be in full operation in 2013. The superb spatial resolution and sensitivity of 
ALMA will allow to uniquely identify the optical/near-IR counterpart of the mm-submm sources.  Furthermore, ALMA will 
directly provide the redshift of the mm-submm sources through the detection of CO rotational transition lines, up to very 
high redshifts. Combining these new data with existing multiwavelength information will finally allow us to complete the
census of supermassive black hole growth since the epoch of cosmic re-ionization.

\section*{Acknowledgements}

Support for the work of ET was provided by the National Aeronautics and Space Administration through Chandra/Einstein Post-doctoral Fellowship 
Award Numbers PF8-90055 and PF9-00069 respectively issued by the Chandra X-ray Observatory Center, which is operated by the Smithsonian 
Astrophysical Observatory (SAO) for and on behalf of the National Aeronautics Space Administration under contract NAS8-03060. Additional support
for this work was provided by NASA through Chandra Award SP1-12005X. ET received partial support from Center of Excellence in Astrophysics and 
Associated Technologies (PFB 06). CMU acknowledges  support from NSF grants AST-0407295, AST-0449678, AST-0807570 and Yale University.

\end{document}